\documentclass[
reprint,
superscriptaddress,
showpacs,
preprintnumbers,
nofootinbib,
nobibnotes,
amsmath,
amssymb, 
aps,
prd,
floatfix
]{revtex4-2}

\pdfoutput=1
\usepackage{amsmath}
\usepackage{amsfonts}
\usepackage{amssymb}
\usepackage{mathrsfs}
\usepackage{color}
\usepackage{slashed}
\usepackage{physics}
\usepackage{cancel}
\usepackage{graphicx}
\usepackage{xcolor}
\usepackage[colorlinks=true,allcolors=blue]{hyperref}
\usepackage{multirow}
\usepackage{upgreek}
\usepackage[capitalise]{cleveref}
\usepackage{xspace}
\usepackage{lipsum}
\usepackage[colorinlistoftodos]{todonotes}

\usepackage{siunitx}

\DeclareSIUnit \s {\second}
\DeclareSIUnit \ns {\nano\second}
\DeclareSIUnit \mus {\micro\second}
\DeclareSIUnit \ms {\milli\second}
\DeclareSIUnit \MB {\mega\byte}
\DeclareSIUnit \GB {\giga\byte}
\DeclareSIUnit \TB {\tera\byte}
\DeclareSIUnit \PB {\peta\byte}
\DeclareSIUnit \Mbps {\mega\bit/\s}
\DeclareSIUnit \Gbps {\giga\bit/\s}
\DeclareSIUnit \Tbps {\tera\bit/\s}
\DeclareSIUnit \Pbps {\peta\bit/\s}
\DeclareSIUnit \kton {\kilo\tonne} 
\DeclareSIUnit \kt {\kilo\tonne}
\DeclareSIUnit \kty {\kilo\tonne-\year}
\DeclareSIUnit \Mt {\mega\tonne}
\DeclareSIUnit \eV {\electronvolt}
\DeclareSIUnit \keV {\kilo\electronvolt}
\DeclareSIUnit \MeV {\mega\electronvolt}
\DeclareSIUnit \GeV {\giga\electronvolt}
\DeclareSIUnit \TeV {\tera\electronvolt}
\DeclareSIUnit \PeV {\peta\electronvolt}
\DeclareSIUnit \EeV {\exa\electronvolt}
\DeclareSIUnit \m {\meter}
\DeclareSIUnit \cm {\centi\meter}
\DeclareSIUnit \nm {\nano\meter}
\DeclareSIUnit \in {\inchcommand}
\DeclareSIUnit \km {\kilo\meter}
\DeclareSIUnit \kV {\kilo\volt}
\DeclareSIUnit \kW {\kilo\watt}
\DeclareSIUnit \MW {\mega\watt}
\DeclareSIUnit \MHz {\mega\hertz}
\DeclareSIUnit \mrad {\milli\radian}
\DeclareSIUnit \year {years}
\DeclareSIUnit \POT {POT}
\DeclareSIUnit \sig {$\sigma$}
\DeclareSIUnit\parsec{pc}
\DeclareSIUnit\lightyear{ly}
\DeclareSIUnit\foot{ft}
\DeclareSIUnit\ft{ft}
\DeclareSIUnit \ppb{ppb}
\DeclareSIUnit \ppt{ppt}
\DeclareSIUnit \samples{S}
\DeclareSIUnit \pe{PE}
\DeclareSIUnit \GeVmwe{GeV/mwe}
\DeclareSIUnit \mwe{mwe}

\newcommand{\enu}{\E_\enu}

\usepackage{isotope}

\newcommand{\refref}[1]{Ref.~\cite{#1}}

\def\minerva{MINERvA\xspace}

\newcounter{CommentCount}
\definecolor{MH}{rgb}{0.0,0.6,9}
\definecolor{palatinate}{rgb}{0.494, 0.192, 0.482}

\renewcommand{\phi}{\varphi}

\begin{document}

\title{Dipole-Coupled Neutrissimo Explanations of the MiniBooNE Excess Including Constraints from \minerva Data}

\author{N.W.~Kamp}
\affiliation{Dept.~of Physics, Massachusetts Institute of Technology, Cambridge, MA 02139, USA}

\author{M.~Hostert}
\affiliation{University of Minnesota, Minneapolis, MN 55455, USA}
\affiliation{Perimeter Institute for Theoretical Physics, Waterloo, ON N2J 2W9, Canada}

\author{A.~Schneider}
\affiliation{Dept.~of Physics, Massachusetts Institute of Technology, Cambridge, MA 02139, USA}

\author{S.~Vergani}
\affiliation{University of Cambridge, Cambridge CB3 0HE, United Kingdom}

\author{C.A.~Arg{\"u}elles}
\affiliation{Dept.~of Physics, Harvard University, Cambridge, MA 02138, USA}

\author{J.M.~Conrad}
\affiliation{Dept.~of Physics, Massachusetts Institute of Technology, Cambridge, MA 02139, USA}

\author{M.H.~Shaevitz}
\affiliation{Dept.~of Physics, Columbia University, New York, NY, 10027, USA}

\author{M.A.~Uchida}
\affiliation{University of Cambridge, Cambridge CB3 0HE, United Kingdom}

\date{\today}

\begin{abstract}
We revisit models of heavy neutral leptons (neutrissimos) with transition magnetic moments as explanations of the $4.8\sigma$ excess of electron-like events at MiniBooNE.
We first re-examine the preferred regions in the model parameter space to explain MiniBooNE, considering also potential contributions from oscillations due to an eV-scale sterile neutrino.
We then derive constraints on the model using neutrino-electron elastic scattering data from \minerva.
To carry out these analyses, we have developed a detailed Monte Carlo simulation of neutrissimo interactions within the MiniBooNE and \minerva detectors using the \texttt{LeptonInjector} framework.
This simulation allows for a significantly more robust evaluation of the neutrissimo model compared to previous studies in the literature--a necessary step in order to begin making definitive statements about beyond the Standard Model explanations of the MiniBooNE excess.
We find that \minerva rules out a large region of parameter space, but allowed solutions exist at the $2\sigma$ confidence level. 
A dedicated \minerva analysis would likely be able to probe the entire region of preference of MiniBooNE in this model.
\end{abstract}

\maketitle

\section{Introduction} \label{sec:intro}
Past, current, and future neutrino experiments offer some of the most promising avenues for observing physics beyond the Standard Model (BSM).
The discovery of neutrino oscillations, and thus nonzero neutrino mass, is itself an indication of BSM physics~\cite{SNO:2002hgz,Super-Kamiokande:2005mbp}, spurring a decades-long oscillation experimental program spanning orders of magnitude in energy and length scales~\cite{deSalas:2020pgw,Esteban:2020cvm,Capozzi:2021fjo}.
A nearly-consistent three-neutrino mixing paradigm has emerged from this program; however, anomalous results have also been observed.
Two striking examples are the excess of inverse-beta-decay events at the LSND detector at the Los Alamos Neutrino Science Center~\cite{LSND:2001aii} and the excess of electron-like events observed by the MiniBooNE (MB) experiment at the Fermilab Booster Neutrino Beam (BNB)~\cite{Aguilar-Arevalo:2020nvw}.
Determining the nature of these excesses is an active frontier in neutrino physics~\cite{Acero:2022wqg}
Historically, both excesses have been interpreted within the context of a $3+1$ model, in which one introduces an eV-scale sterile neutrino facilitating short-baseline $\nu_\mu (\overline{\nu}_\mu) \to \nu_e (\overline{\nu}_e)$ oscillations.
However, models beyond the vanilla $3+1$ scenario may better accommodate these anomalies within the global experimental landscape, including cosmology and other neutrino oscillation experiments.

One such model considers MeV-scale heavy neutral leptons (HNLs), or ``neutrissimos''~\cite{Loinaz:2004qc,NuSOnG:2008weg}, with transition magnetic moments, popularly known as the dipole portal to HNLs~\cite{Gninenko:2009ks,Gninenko:2010pr,Dib:2011jh,Gninenko:2012rw,Masip:2012ke,Radionov:2013mca,Ballett:2016opr,Magill:2018jla,Balantekin:2018ukw,Balaji:2019fxd,Balaji:2020oig,Fischer:2019fbw,Vergani:2021tgc,Alvarez-Ruso:2021dna}.
Most recently, \refref{Vergani:2021tgc} showed that the MB anomaly could be fitted with such a dipole portal extended to include a mass-mixed eV-scale sterile neutrino.
The advantages of this model are twofold: (1) it provides a better description of the low-energy, forward-angle part of the MB excess compared to the $3+1$ model alone, and (2) reducing the oscillation-based contribution to the MB excess alleviates tension in global fits to the $3+1$ model while retaining an explanation to the LSND anomaly. 

For the main result of this paper, we derive constraints on the dipole portal coupling using existing experimental results from the \minerva collaboration.
Specifically, \minerva has performed measurements of the neutrino-electron elastic scattering ($\nu-e$) rate for the purpose of constraining the Neutrino Main Injector (NuMI) low-energy (LE) and medium-energy (ME) neutrino fluxes~\cite{Park:2015eqa,Valencia:2019mkf,MINERvA:2022vmb}.
Photons produced in the decay of the dipole-coupled neutrino would mimic the single electromagnetic shower morphology of electron-scattering (ES) events and would therefore enter as a photon-like (large $dE/dx$) background in the \minerva analysis.
In addition to deriving new constraints from \minerva, we improve upon the MiniBooNE dipole model analysis performed in Ref.~\cite{Vergani:2021tgc}.

The analyses described in this article have been carried out using a novel simulation developed within the \texttt{LeptonInjector} framework~\cite{IceCube:2020tcq}.
This tool allows a more robust description of neutrissimo interactions in detector subsystems as well as the position and kinematics of observable final state particles.
This is an important improvement over previous treatments in the literature, as an accurate simulation of exotic BSM physics scenarios in different neutrino detectors will be vital in determining the nature of the MiniBooNE excess.

The rest of this article is organized as follows.
In section~\ref{sec:model}, we review the dipole-portal sterile neutrino model in more detail.
In section~\ref{sec:simulation}, we introduce the novel \texttt{LeptonInjector}-based simulation developed for the studies presented here.
In section~\ref{sec:mb}, we refine the preferred regions in dipole parameter space which explain the energy and angular distributions of the MiniBooNE excess.
In section~\ref{sec:mvresults}, we calculate constraints in dipole parameter space derived from the \minerva ES analysis.
In section~\ref{sec:discussion}, we discuss existing and projected constraints on the dipole model from current and future neutrino experiments.
Finally, in section~\ref{sec:conclusion} we discuss the outlook of this mixed model of oscillation and decay as a solution to the MiniBooNE anomaly in light of our derived \minerva constraints.

\section{The Dipole Portal} \label{sec:model}
\begin{figure}[t]
\begin{center}
\includegraphics[width=\columnwidth]{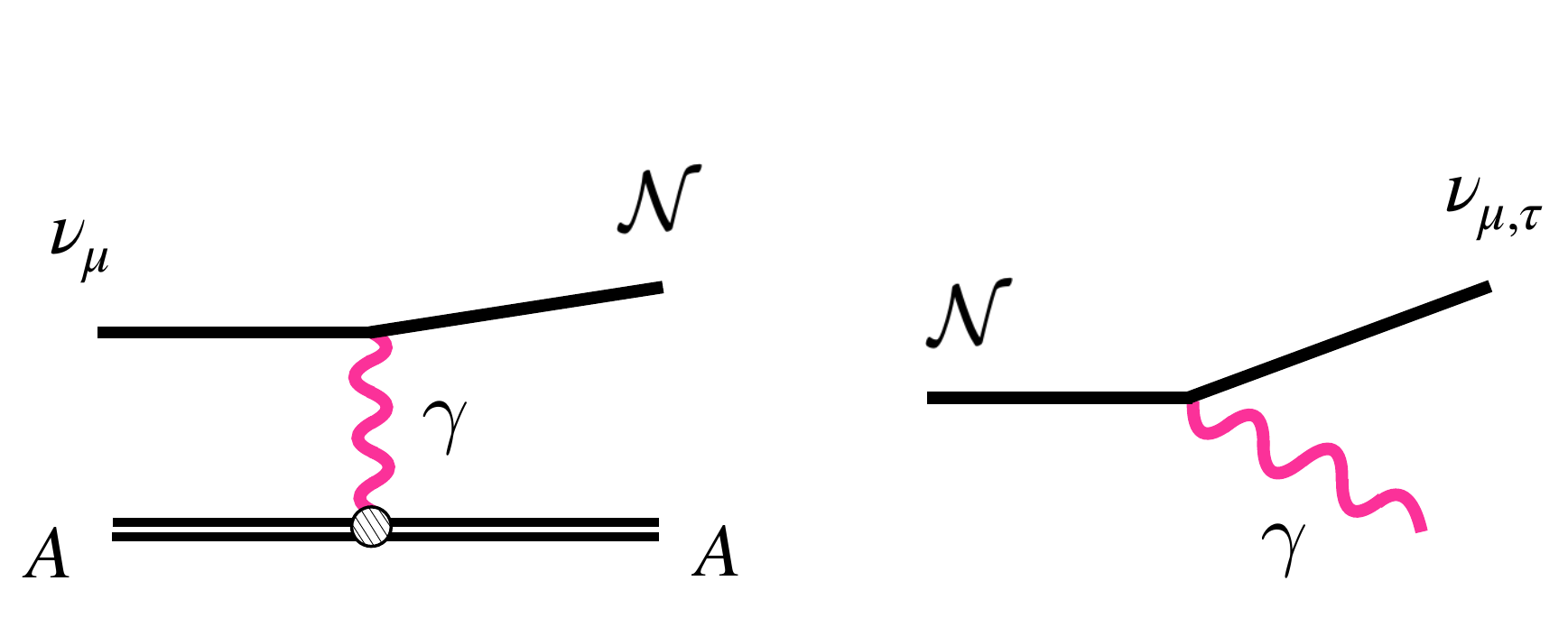}
\end{center}
\vspace{-0.6cm}
\caption{$\mathcal{N}$ production from $\nu$ upscattering (left) and $\mathcal{N}$ decay (right).}
\label{fig:feynman}
\end{figure}

We introduce a right-handed neutrino $\mathcal{N}$ that couples to the left-handed neutrino fields via a transition magnetic moment.
At the effective operator level, we have the dimension-six dipole operators,
\begin{align}\label{eq:dim6_dipoles}
    \mathscr{L} &\supset \frac{1}{\Lambda^2} \overline{L_\alpha}\widetilde{H} \sigma^{\mu \nu}  \mathcal{N}_R\left( C_B^\alpha \, B_{\mu \nu} + C_W^\alpha\,W^a_{\mu \nu}\sigma_a\right) + \text{ h.c.},
\end{align}
where $\alpha$ is a flavor index, $\widetilde{H} = i\sigma_2 H^*$, $C_B^\alpha$, and $C_W^\alpha$ are Wilson coefficients and $\Lambda$ is the new physics scale.
After electroweak symmetry breaking, the dipole operator gives rise to the electromagnetic transition magnetic moment of neutrinos,
\begin{align}\label{eq:dipole}
     \mathscr{L} &\supset  d_{\alpha \mathcal{N}}\, \overline{\nu_\alpha} \sigma_{\mu\nu}  F^{\mu \nu} \mathcal{N}_R  + \text{ h.c.},
\end{align}
where $\nu_\alpha$ corresponds to the neutrino Weak eigenstates and $F^{\mu \nu}$ to the electromagnetic field strength.
The dipole parameter is defined as $d_{\alpha \mathcal{N}} =  (v_{h}/\sqrt{2})(C_W^\alpha C_B^\alpha +C_W^\alpha s_W)/\Lambda^2$, where $v_h$ is the vacuum expectation value of the SM Higgs. 
The other transition moments mediated by the $W$ and $Z$ bosons will also be present, but their low-energy effects are further suppressed by $G_F$, and therefore negligible in our region of interest.

The upscattering signature we are interested in at MiniBooNE is initiated by muon neutrinos and antineutrinos, so in this work, we always consider the second-generation coupling $d_{\mu\mathcal{N}}$. 
In UV-completions of this operator from one-loop diagrams, the size of the dipole coupling is typically proportional to $d_{\alpha\mathcal{N}} \sim (1/16 \pi^2) (m_\beta/m_X^2)$, where $m_\beta$ is the mass of some charged particle and $X$ is some heavy, charged scalar, for example.
Under the assumption of flavor-conserving interactions between SM neutrinos and the new physics, one would take $\beta = \alpha$ and conclude that the transition magnetic moment of the third-generation neutrinos is much larger.
While this need not necessarily be the case, it still provides enough motivation for us to consider the third-generation coupling 
$d_{\tau\mathcal{N}}$, including the case
\begin{equation}
\frac{d_{\tau\mathcal{N}}}{d_{\mu\mathcal{N}}} = \frac{m_\tau}{m_\mu},    
\end{equation}
and neglecting the first-generation couplings altogether.

We note that large neutrino magnetic moments typically imply large Dirac masses for light neutrinos, $m_D \overline{\nu_L} \mathcal{N}_R$~\cite{Fujikawa:1980yx,Pal:1981rm,Shrock:1982sc}.
In models with heavy neutrinos, this presents two challenges: i) neutrino masses, schematically given by $m_\nu \sim m_D^2/M$, may be too large, and ii) $m_D$ will generate large mixing between active and heavy neutrinos, $U_{\alpha \mathcal{N}} \sim m_D/M$.
Here, $M$ stands for the heavy neutrino mass scale.
The first challenge is easily overcome in models like the inverse-seesaw where lepton number is approximately conserved and right-handed neutrino fields combine into pseudo-Dirac $\mathcal{N}$ particles.
In that case, $m_\nu$ is controlled by the mass splitting between the Majorana neutrinos, which may be parametrically small.
This is also the case preferred by the short-baseline phenomenology discussed below, since the single-photons produced in the decays of Dirac HNLs are less forward.
The second challenge, however, is not so easy to overcome. The mixing between active and heavy neutrinos remains large even in inverse-seesaw models and the parameter space in that case is strongly constrained by laboratory limits on $U_{\alpha \mathcal{N}}$.
For instance, decay-in-flight signatures at neutrino experiments, where $\mathcal{N}$ is copiously produced in meson decays at the target, can set limits as strong as $|U_{\alpha \mathcal{N}}|^2 < \mathcal{O}(10^{-11})$~\cite{Arguelles:2021dqn}.

There are several models in the literature that can suppress the Dirac mass in comparison with the magnetic moment of neutrinos~\cite{Voloshin:1987qy,Barbieri:1988fh,Babu:1989px,Babu:1989wn,Leurer:1989hx,Lindner:2017uvt,Babu:2020ivd}, but only to a certain extent and not without fine tuning. 
We proceed assuming that $m_D$ is sufficiently small so as not to impact the phenomenology, but note that depending on the amount of fine tuning, bounds on the mixing angles $|U_{\alpha \mathcal{N}}|$ would also need to be considered.

In performing a fit to MiniBooNE data, we also include a sterile neutrino, $\nu_s$, which mixes with light neutrinos but does not have transition magnetic moments with active neutrinos.
This sterile will be responsible for short-baseline oscillations with $\Delta m_{41}^2$ of $\mathcal{O}(1$~eV$^2)$.
We provide two examples, the case of a global fit to short-baseline data excluding MiniBooNE~\cite{Vergani:2021tgc} and the case of a joint fit to only MiniBooNE and the recent MicroBooNE CCQE-like analysis~\cite{MiniBooNE:2022emn}.
It is also important that $\nu_s$ and the heaviest neutrino do not mix, as otherwise $\mathcal{N}_R$ would also mix with light neutrinos via $\nu_s$.
In summary, our spectrum is defined as
\begin{equation}
    \nu_i \sim \sum_{\alpha=\{e,\mu,\tau,s\}} U_{\alpha i} \nu_\alpha, \text{ for $i\leq 4$ and  } \nu_5 \equiv \mathcal{N}
\end{equation}
where the approximate symbol means that any additional term contains very small mixing elements.
In addition, this mixing would result in a corresponding dipole coupling $d_{\alpha s} \overline{\nu}_\alpha \sigma_{\mu \nu} F^{\mu \nu} \nu_s $, which is strongly constrained by Big-Bang Nucleosynthesis and stellar cooling.
While the former may be modified {\it à la} secret interactions, the latter is significantly more robust and constrains $d_{\mu s} < 2\times 10^{-12} \mu_B$~\cite{Brdar:2020quo}.

The relevant interactions for this work are shown in Fig.~\ref{fig:feynman}, which include Primakoff upscattering off of a nuclear target $\nu A \to \mathcal{N} A$ (left) and the radiative decay of the heavier neutrino $\mathcal{N} \to \nu_\alpha \gamma$ (right). 
Due to the photon propagator, the scattering process is dominated by coherent exchange with the nucleus, except at the largest $m_{\mathcal{N}}$ values. 
We include helicity-flipping upscattering, where the helicity of $\mathcal{N}$ is opposite to that of $\nu_\alpha$, and helicity-conserving, where they are the same.
The latter is suppressed by the typical energy of the process,  $m_{\mathcal{N}}^2/E^2$, and is only relevant at the largest masses.
To give an example, the cross section for an incoming neutrino with energy $E_{\nu_\alpha} = 1\;{\rm GeV}$ upscattering off of a carbon nucleus into a neutrissimo with mass $m_\mathcal{N} = 50\;{\rm MeV}$ and dipole coupling $d_{\alpha N} = 10^{-6}\;{\rm GeV}^{-1}$ is $\sigma \approx 5 \times 10^{-39}\;{\rm cm}^2$~\cite{Magill:2018jla}. 

The rest frame decay width of the $\mathcal{N}$ is given by the incoherent sum over all outgoing flavors,
\begin{equation}\label{eq:dipole_decay}
\Gamma_{\mathcal{N} \to \nu \gamma} = \left( \sum_{\alpha} |d_{\alpha \mathcal{N}}|^2 \right) \frac{m_\mathcal{N}^3}{4 \pi}.
\end{equation}
Throughout this work, we will assume that either $\mathcal{N}$ or light neutrinos are Dirac particles, such that the differential decay rate is proportional to $(1 \pm \cos{\theta})$~\cite{Balantekin:2018ukw}.
For the region of dipole parameter space preferred by MiniBooNE~\cite{Vergani:2021tgc}, the HNL has a lab frame decay length $L \sim 1-10\;{\rm m}$ for typical \minerva neutrino energies and $L \leq 1\;{\rm m}$ for typical MiniBooNE neutrino energies.
Thus, Primakoff upscattering near or within each detector is the most relevant production mechanism in this region of parameter space.
For smaller HNL masses, decay lengths become longer and upscattering in the dirt along the BNB and NuMI beamlines becomes more important for MiniBooNE and \minerva, respectively.


\section{Simulating Neutrissimos}
\label{sec:simulation}

\begin{figure}[t]
    \centering
    \includegraphics[width=0.49\textwidth]{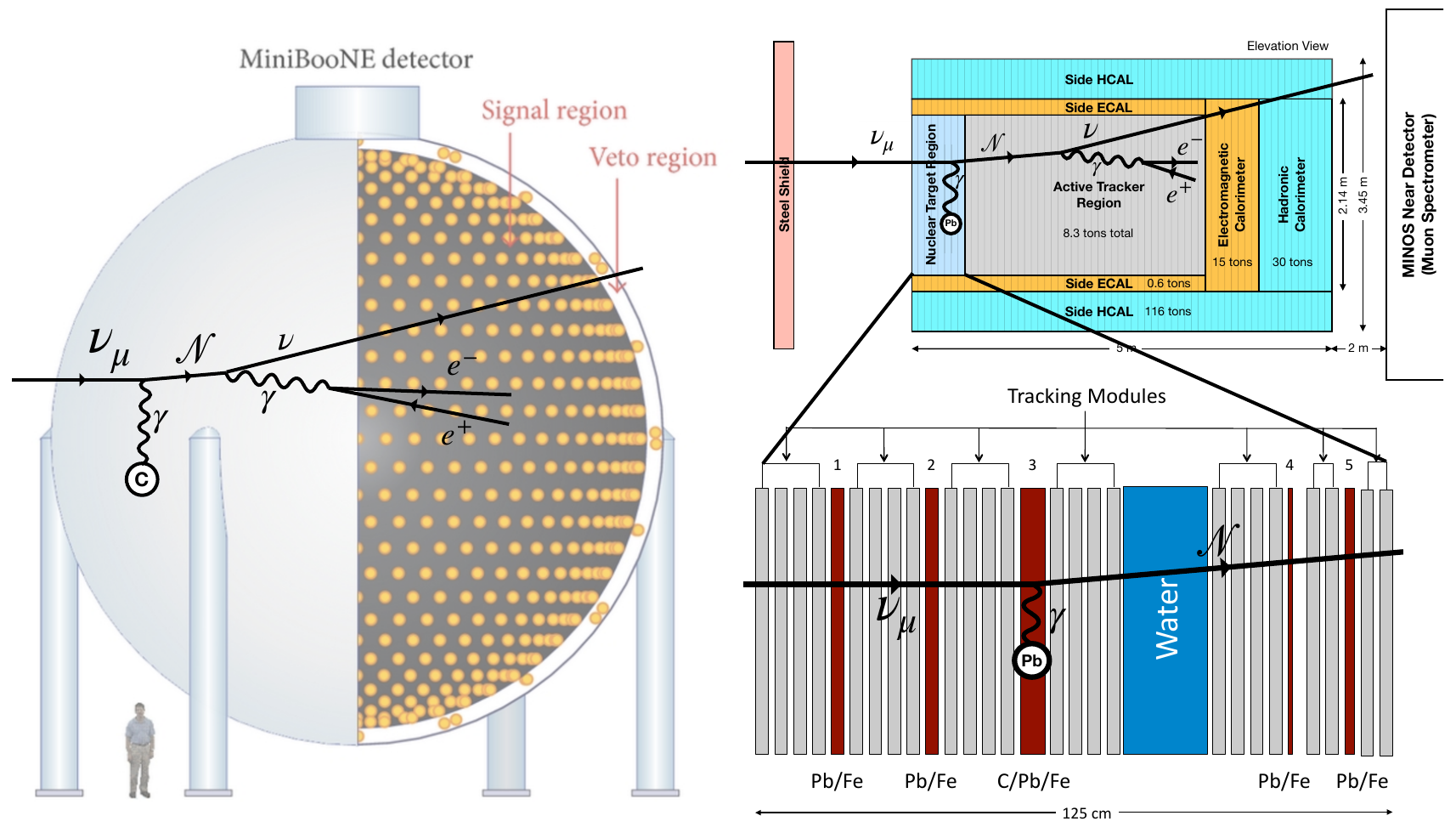}\\
    \vspace{2ex}
    \includegraphics[width=0.49\textwidth]{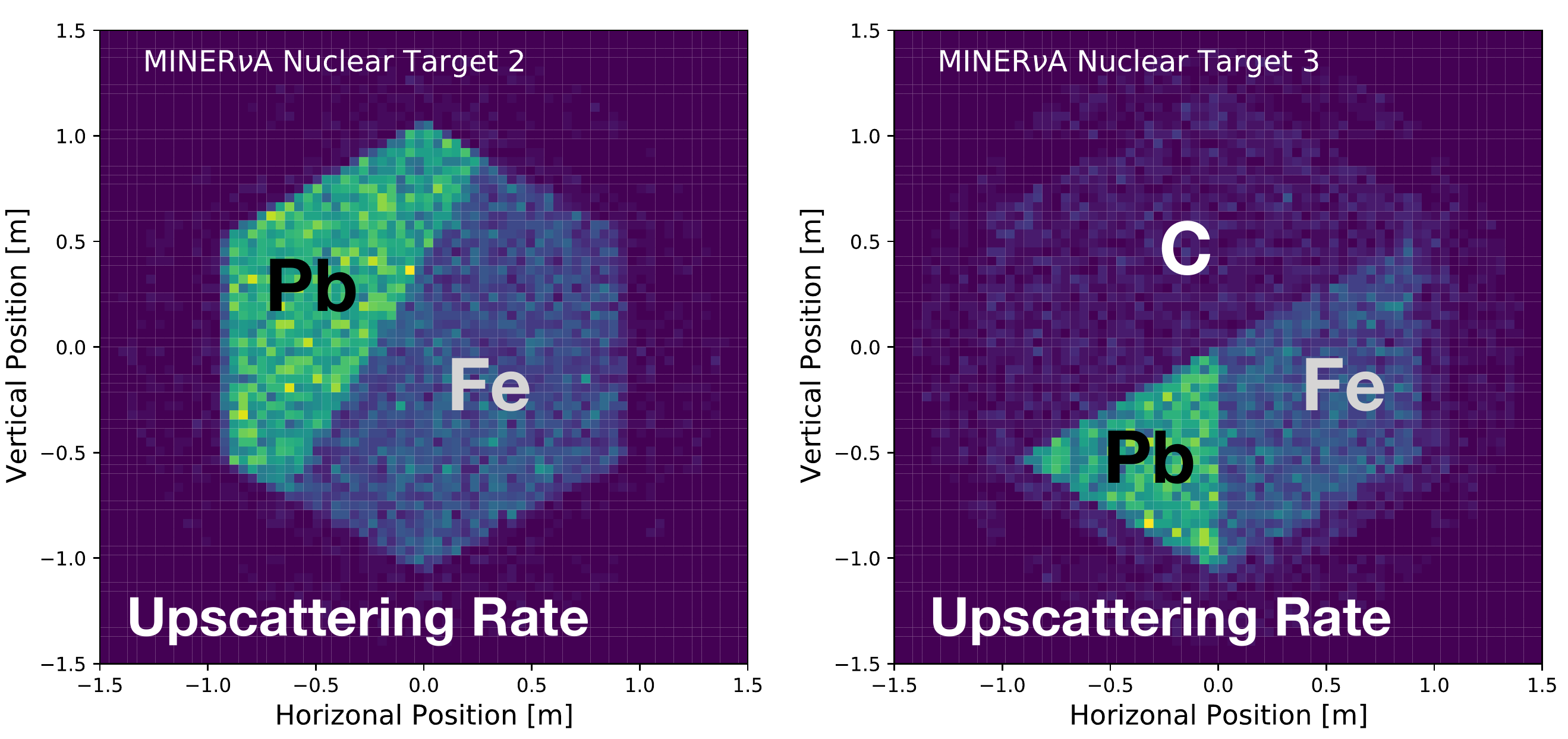}
    \caption{Top) schematic representation of HNL production via upscattering and the subsequent HNL decay within the MiniBooNE (left) and \minerva (right) detectors. Detector images have been adapted from \refref{MiniBooNE:2008paa} and \refref{MINERvA:2013zvz}. Bottom) example upscattering rates within two of the \minerva nuclear targets as simulated using \texttt{LeptonInjector}. The coherent enhancement of the upscattering cross sections leads to a larger rate in the high-$Z$ components.}
    \label{fig:lepinj_example}
\end{figure}

In order to describe neutrissimo interactions in the MiniBooNE and \minerva detectors, we have developed a custom simulation based in the \texttt{LeptonInjector} framework~\cite{IceCube:2020tcq}.
The simulation begins by injecting neutrinos in MiniBooNE and \minerva according the BNB and NuMI beam profiles, respectively.
For each neutrino, a flight path is randomly selected within a cone surrounding the detector.
This path is then used to calculate intersections with different components of the beamline, including bedrock between the target and detector as well as various detector subsystems.
An upscattering location is sampled along this flight path according to the cross section in each traversed material.
The final state kinematics of the produced HNL are sampled according to the $d\sigma/dy$ distribution, where $y \equiv (E_\nu - E_\mathcal{N})/E_\nu$ in the lab frame~\cite{Brdar:2020quo}.
See the \Cref{app:xsecs} for more details on the upscattering cross section used in this analysis.
Next, a decay location is sampled along the flight path of the HNL.
The final state kinematics of decay photon are sampled according to $d\Gamma / d\cos\theta \propto (1-\cos\theta)$.
We trace the flight path of the photon through different materials until it converts to an $e^+ e^-$ pair, which are assumed to reconstruct as a single electromagnetic shower within the MiniBooNE and \minerva detectors.

The above procedure is easily generalizable to other BSM scenarios and detector configurations.
The user needs only to provide (1) the flux of initial state particles (neutrinos in this case), (2) the total and differential cross sections for the relevant processes, and (3) the relevant detector geometry.
The flux and cross section can be input either as analytic expressions or splines; \texttt{LeptonInjector} will interpolate in the latter case.
The detector geometry is set using a simple configuration file in which the user instantiates any number of volumes.
\texttt{LeptonInjector} supports most simple volumes as well as general extruded polygons--for example, the nuclear targets in \minerva shown in \cref{fig:lepinj_example}, which are subsections of hexagonal prisms).
\texttt{LeptonInjector} will then generate user-specified initial state particles, sample their interaction locations, and store the kinematics and weights of the final state particles.
The simulation is set up to be as efficient as possible, such that most generated events create an observable final state within the fiducial volume and are appropriately down-weighted.
In the analysis presented here, we have tracked final state particles until the production of an observable $e^+ e^-$ pair.
This strategy can be adapted for other BSM scenarios which involve multiple interactions before the production of an observable final state.

The simulation performed for the analysis presented here gives a robust estimation of the $e^+ e^-$ event rate within the fiducial volume of each experiment.
Further, it provides an accurate kinematic description of the photons which survive fiducialization (namely, the photon energy and angle with respect to the beamline).
This is vital, as it will be shown that the region in dipole parameter space preferred by the MiniBooNE excess is highly sensitive to the kinematics of the final state photon.
Such an effect has been appreciated by previous studies in the literature~\cite{Gninenko:2009ks,Alvarez-Ruso:2021dna}, but a comprehensive fit over the full dipole parameter space was not attempted until \refref{Vergani:2021tgc}; the simulation described above allows us to refine the fit from \refref{Vergani:2021tgc}.

The simulation is also vital in order to properly assess the ability of \minerva to constrain the dipole solution to the MiniBooNE excess.
By leveraging our simulation, we will show that the reconstruction efficiency for radiative HNL decays in \minerva varies by 3-4 orders of magnitude across the parameter space.
Previous calculations in the literature have assumed a constant 10\% reconstruction efficiency for radiative HNL decays in \minerva~\cite{Brdar:2020tle}, and have thus incorrectly concluded that \minerva data rules out the dipole-portal MiniBooNE solution.
Further, they do not simulate the complex sub-components of the \minerva detector--this is an important step, as upscattering in the high-$Z$ nuclear targets just upstream of the fiducial volume can contribute significantly to the single shower sample for short-lived HNLs.

Thus, the \texttt{LeptonInjector}-based simulation performed for this study is essential to evaluate the status of the dipole-coupled neutrissimos as an explanation for the MiniBooNE excess.
As the community turns toward more exotic BSM explanations of the MiniBooNE excess, it is imperative that these models are evaluated within the context of realistic detector descriptions.
Our simulation framework is the prefect tool for such a task; though the version used in this article was developed specifically to study neutrissimo interactions in MiniBooNE and \minerva, it can be adapted easily to accommodate additional BSM scenarios and neutrino detectors.

\section{Neutrissimos at MiniBooNE} \label{sec:mb}

The MiniBooNE detector uses Cherenkov light to detect final state particles produced in neutrino interactions.
As electrons and photons both produce electromagnetic showers which show up as distorted Cherenkov rings in the detector, the two particles are indistinguishable in MiniBooNE.
Thus, photons from the dipole model could contribute to the MiniBooNE electron-like excess.
In the case of a nonzero effective dipole coupling, $d_{\mu \mathcal{N}}$, muon neutrinos from Fermilab's Booster Neutrino Beam will undergo Primakoff upscattering into HNL states.
This can happen via photon exchange with nuclear targets both within the dirt between the BNB target and the detector and within the CH$_2$ detector volume itself.

\subsection{Simulation}
As described in section~\ref{sec:simulation}, we use \texttt{LeptonInjector} to simulate the production and decay of HNLs in MiniBooNE~\cite{IceCube:2020tcq}.
A schematic depiction of this process is shown in the top panel of Fig.~\ref{fig:lepinj_example}.
Muon neutrinos are injected according to the BNB flux and allowed to upscatter to HNLs along the 541$\;$m baseline between the BNB target and the detector or within the 818-ton CH$_2$ detector itself.
We describe the MiniBooNE detector as a sphere of CH$_2$ with a total radius of 6.1\;m and a fiducial radius of 5\;m.
The MiniBooNE detector sits within a sphere of air with a radius of 9\;m, meant to represent the detector hall.
To simulate the BNB, muon neutrinos generated in \texttt{LeptonInjector} propagate through 541\;m of dirt before reaching the MiniBooNE detector.
The neutrino upscattering and subsequent HNL decay are simulated according to the proceudre outlined in section~\ref{sec:simulation}. 
We parameterize the post-fiducialization photon detection efficiency as a linearly decreasing function of the true photon kinetic energy~\cite{Vergani:2021tgc}.
We also impose a reconstruction threshold on the photon kinetic energy of 140\;MeV~\cite{MiniBooNE:2012maf}.
We also independently smear the reconstructed visible energy and scattering angle of each photon according to the resolution of each as a function of true photon kinetic energy.
The energy (angular) resolution comes from a power-law (quadratic) fit to simulated single electromagnetic shower events in MiniBooNE~\cite{MBres}, with typical values of 10\% (3$^\circ$) for photons from this model, consistent with figures reported by the MiniBooNE collaboration~\cite{Shaevitz:2008zza}.

\subsection{Analysis Methodology}

\begin{figure}[t]
    \centering
    \includegraphics[width=0.45\textwidth]{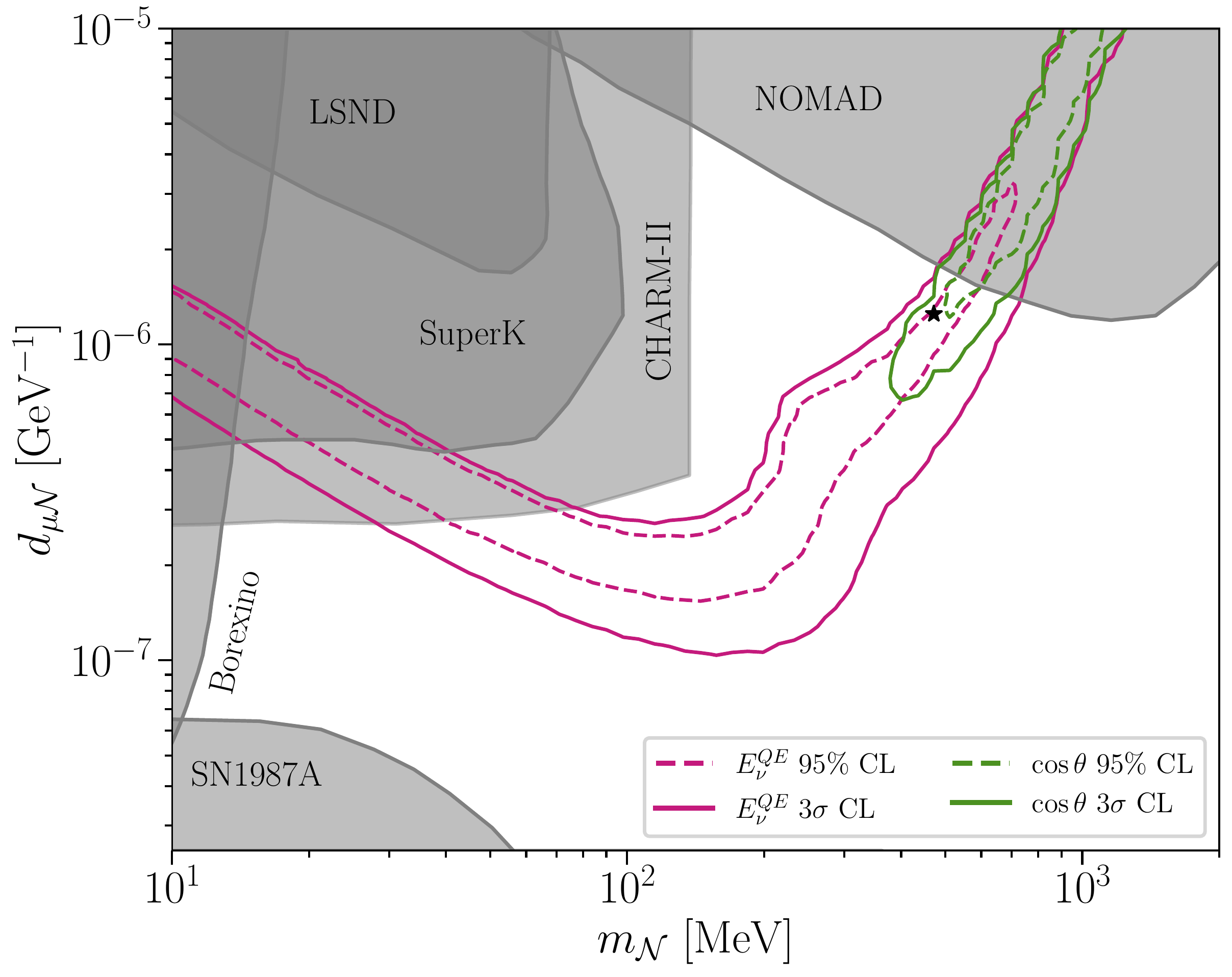}
    \caption{The 2$\sigma$ and 3$\sigma$ CL preferred regions to explain the MiniBooNE anomaly in mass-coupling parameter space for a dipole-coupled heavy neutral lepton. The pink (green) curves correspond to results from fitting the $E_\nu^{\rm QE}$ ($\cos \theta$) distribution. Dipole model fits are performed after subtracting the oscillation component from a global fit to a $3+1$ model excluding MiniBooNE data~\cite{Vergani:2021tgc}. In general, the energy and angular distributions prefer different regions of parameter space, though overlap exists at the $2-3\sigma$ level. \label{fig:miniboone_result}}
\end{figure}

\begin{figure}[t]
    \centering
    \includegraphics[width=0.45\textwidth]{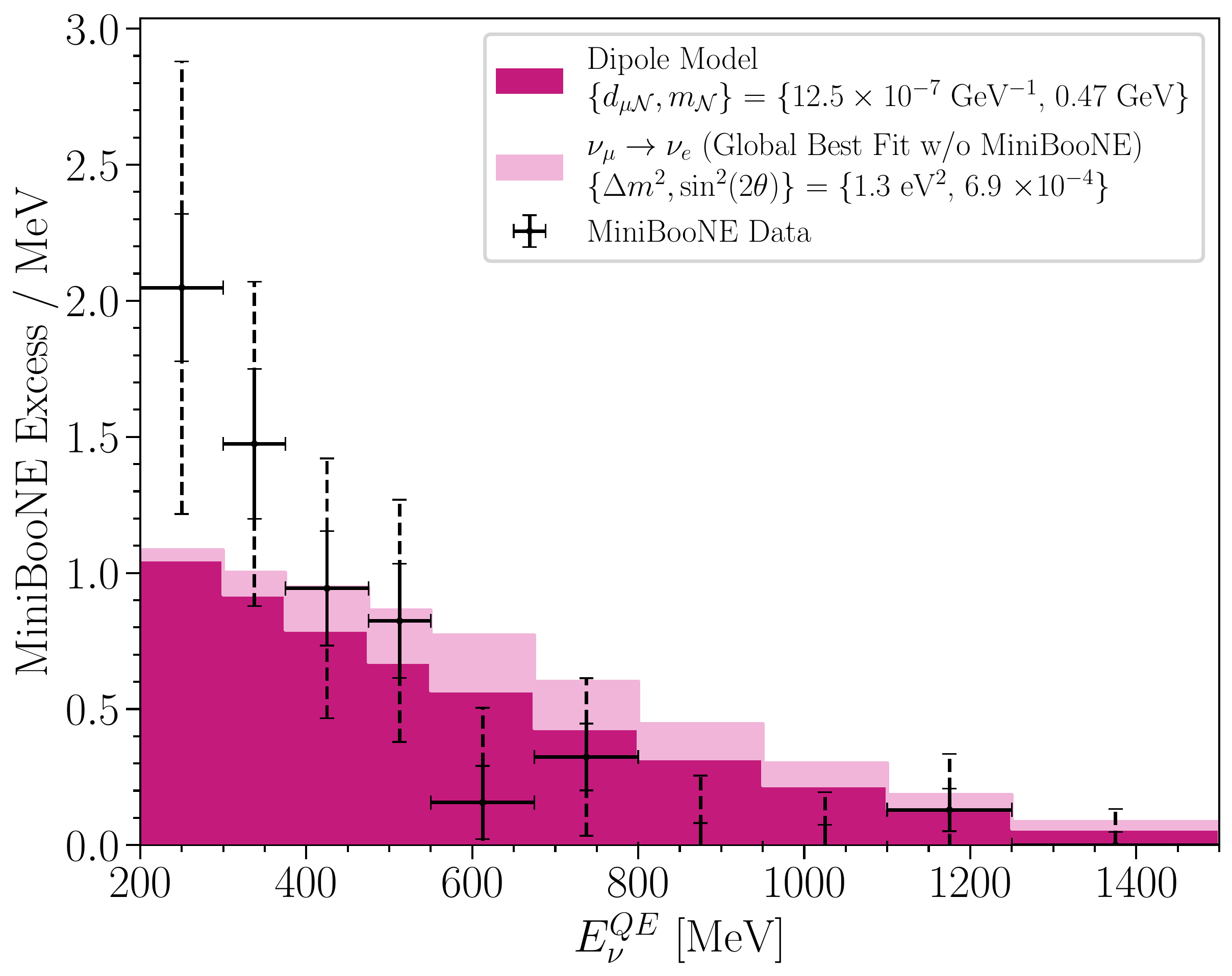}
    \includegraphics[width=0.45\textwidth]{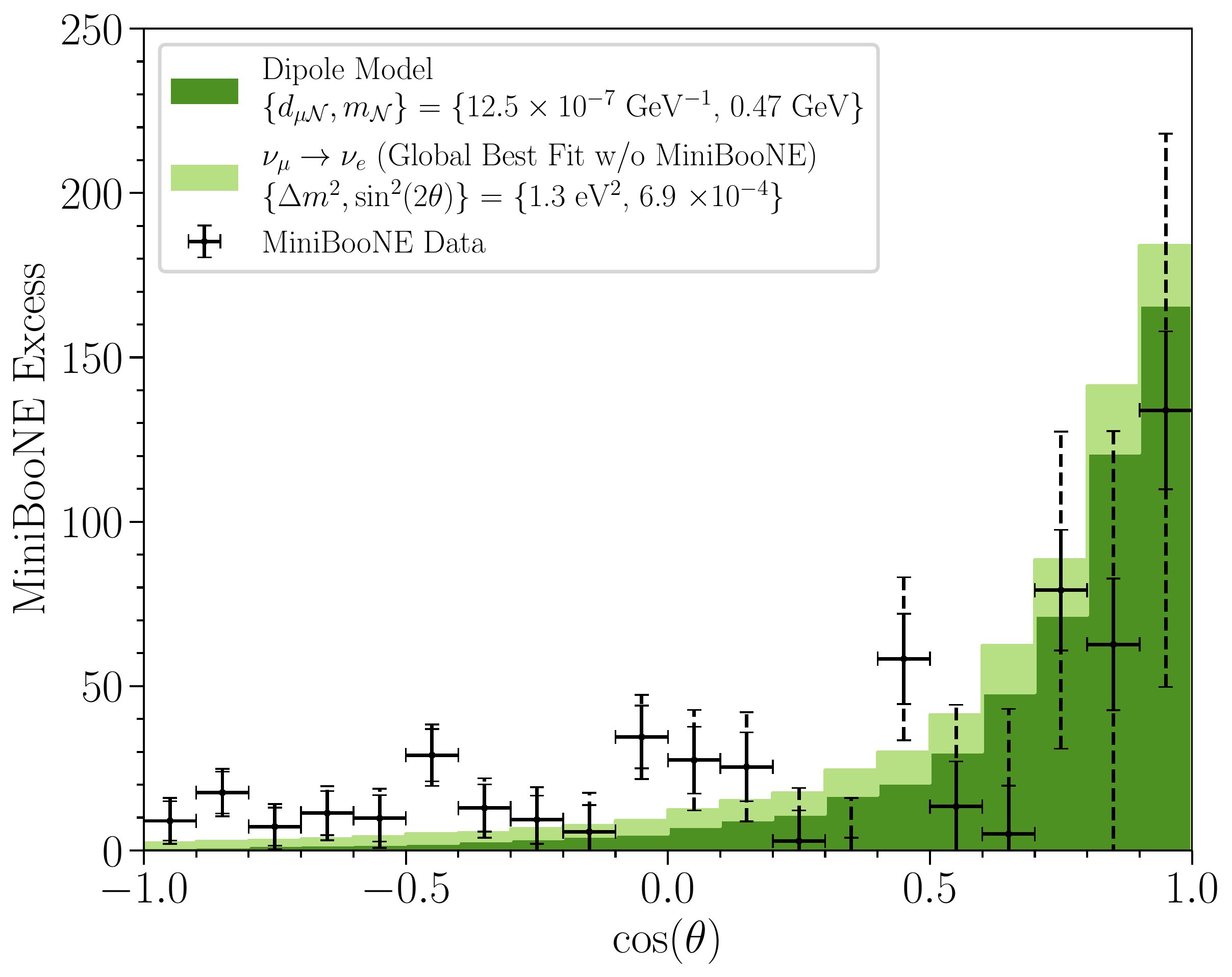}
    \caption{The $E_\nu^{\rm QE}$ (top) and $\cos \theta$ distributions at the example dipole model hypothesis indicated by the black star in Fig.~\ref{fig:miniboone_result}. The darker contribution in each stacked histogram corresponds to the dipole model prediction, while the lighter contribution corresponds to the oscillation contribution. The background-subtracted MiniBooNE excess is indicated by the black data points, with solid and dashed error bars indicating statistical and statistical+systematic errors, respectively.\label{fig:miniboone_dists}}
\end{figure}

The dipole model has previously been studied as a potential solution to the MiniBooNE anomaly~\cite{Gninenko:2009ks,Gninenko:2010pr,Dib:2011jh,Gninenko:2012rw,Masip:2012ke,Radionov:2013mca,Ballett:2016opr,Magill:2018jla,Fischer:2019fbw,Vergani:2021tgc,Alvarez-Ruso:2021dna}.
Specifically, we expand upon the study performed in Ref.~\cite{Vergani:2021tgc}, which examined a mixed model consisting of an eV-scale $\nu_4$ facilitating short-baseline $\nu_\mu \to \nu_e$ oscillations as well as an MeV-scale HNL decaying to a photon via the dipole portal mechanism.
Ref.~\cite{Vergani:2021tgc} found a preference for a dipole-coupled HNL with $m_\mathcal{N} \sim 400\;{\rm MeV}$ and $d_{\mu \mathcal{N}} \sim 3\times10^{-7}\;{\rm GeV}^{-1}$. 

We make a number of improvements to this analysis.
First, we make use of a more robust nuclear electromagnetic form factor, implementing a data-driven Fourier-Bessel function parametrization~\cite{Fricke:1995zz,DeVries:1987atn,DeJager:1974liz} with the data files made available in Ref.~\cite{VT_NDT}. This should be compared with the simpler dipole parameterization, which overestimates the differential cross sections at larger momentum exchange (see \Cref{app:xsecs} for more details).
This has an impact on the allowed regions at larger heavy neutrino masses, as the form factor used in this study drops off much more quickly at larger $Q^2$.
It also reduces the contribution from the dipole model at large scattering angles, thereby making it difficult to explain the back-scattered lepton angular distribution of the MiniBooNE excess.
This effect has been pointed out in previous studies of the dipole model in MiniBooNE~\cite{Radionov:2013mca}.
In this work, we perform a more detailed analysis of the dipole parameter space to determine whether solutions exist which can accommodate both the energy and angular distributions of the excess.
The statistical treatment for each distribution has been improved--we now consider correlated systematic errors in the reconstructed $E_\nu^{\rm QE}$ distribution from the provided covariance matrix (after constraining with the covariance matrix for MiniBooNE's $\nu_\mu$ dataset).
We also introduce an uncorrelated systematic error of 13\% in the $\cos \theta$ distribution, consistent with that in the $E_\nu^{\rm QE}$ distribution.

We perform fits to the excess only in neutrino-mode data, as MiniBooNE has collected about an order of magnitude more events in this beam configuration compared to their antineutrino-mode data. 
We use the simulated photon events from the above procedure to perform two different spectral analyses across dipole parameter space: one in the $E_\nu^{\rm QE}$ distribution and one in the $\cos \theta$ distribution.
In both cases, we calculate a $\chi^2$ test statistic comparing the dipole model prediction to the remaining excess after subtracting off the oscillation contribution from the MiniBooNE-less global fit reported in Ref.~\cite{Vergani:2021tgc}.
In the $E_\nu^{\rm QE}$ fit, we use the electron-like channel fractional covariance matrix provided by the MiniBooNE collaboration after constraining with the covariance matrix in the muon-like channel. 
No systematic errors are provided for the $\cos \theta$ distribution; therefore, as mentioned above, we consider an uncorrelated fractional systematic error of 13\% in each bin of the $\cos \theta$ prediction, consistent with the level in the $E_\nu^{\rm QE}$ channel.
Confidence regions are drawn using a $\Delta \chi^2$ test statistic, assuming Wilks' theorem with two degrees of freedom~\cite{10.1214/aoms/1177732360}. 

\subsection{Results}

\begin{figure}[t]
    \centering
    \includegraphics[width=0.45\textwidth]{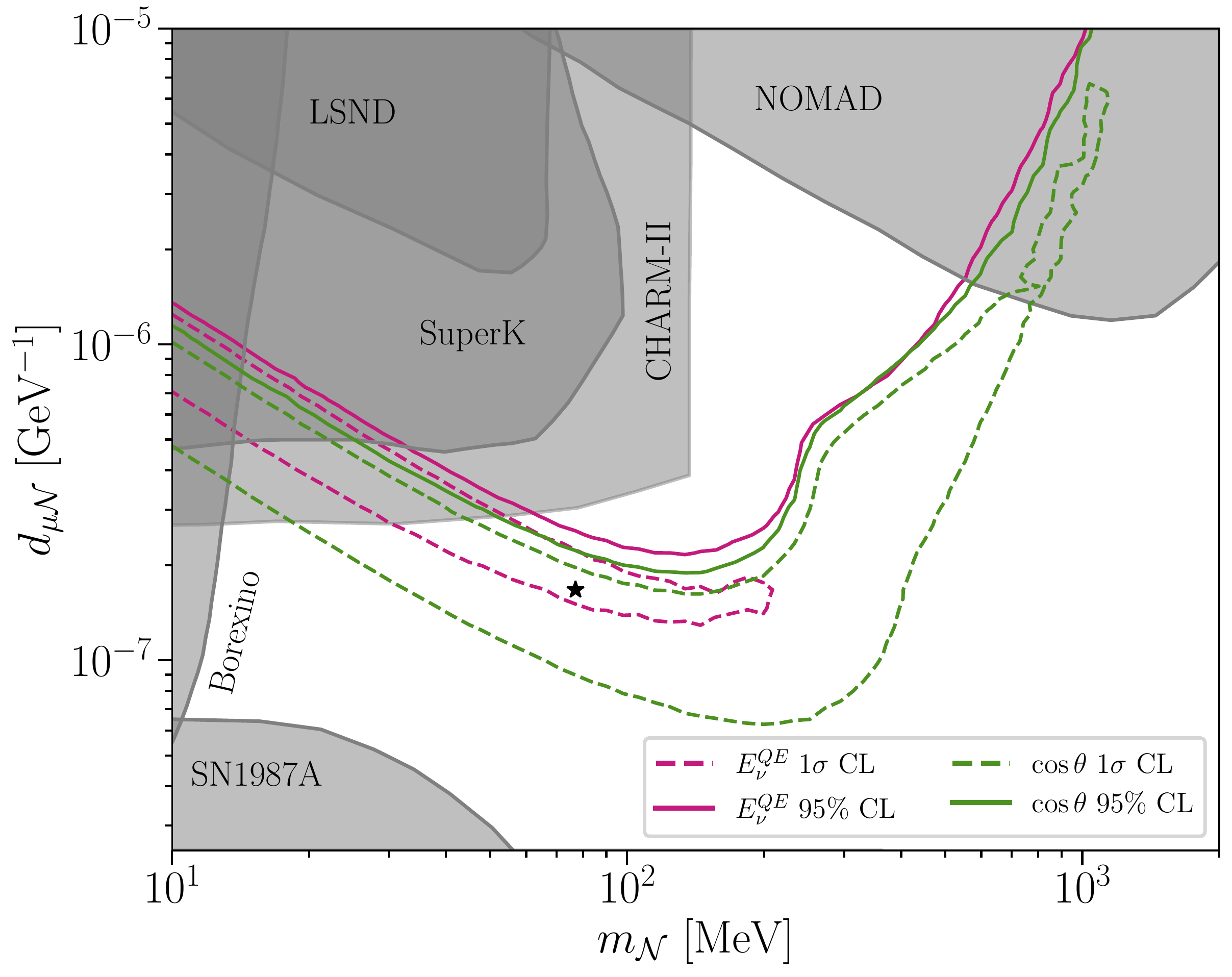}
    \caption{1$\sigma$ CL preferred and 2$\sigma$ CL allowed regions with regard to the MiniBooNE anomaly in mass-coupling parameter space for a dipole-coupled heavy neutral lepton. The pink (green) curves correspond to results from fitting the $E_\nu^{\rm QE}$ ($\cos \theta$) distribution. Dipole model fits are performed after subtracting the oscillation component from the combined MiniBooNE + MicroBooNE CCQE-like $3+1$ fit~\cite{MiniBooNE:2022emn}. Mild preference for a dipole-coupled heavy neutrino with $m_\mathcal{N} \lesssim 100\;{\rm MeV}$ is found at the 1$\sigma$ level. \label{fig:miniboone_result_jointosc}}
\end{figure}

\begin{figure}[t]
    \centering
    \includegraphics[width=0.45\textwidth]{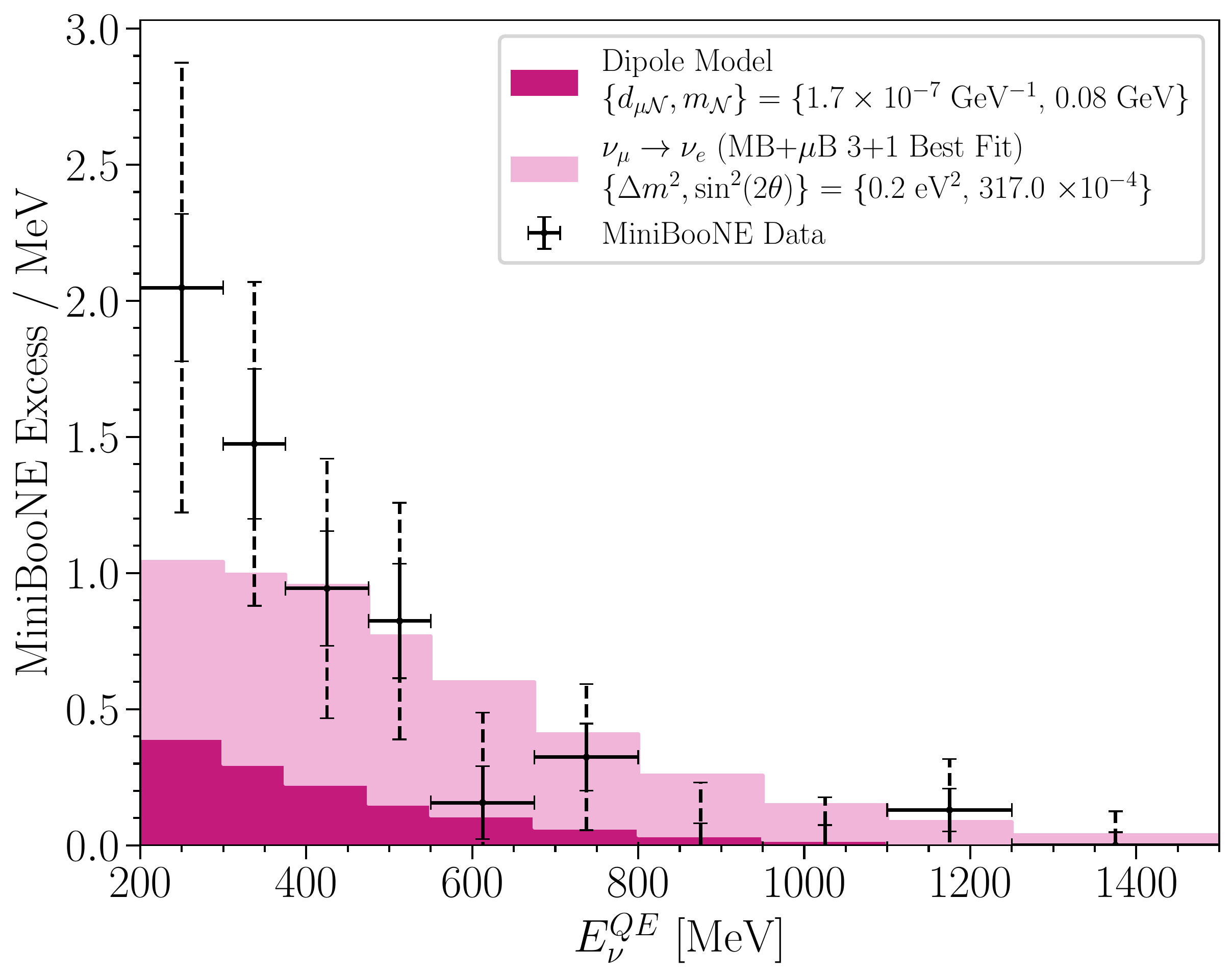}
    \includegraphics[width=0.45\textwidth]{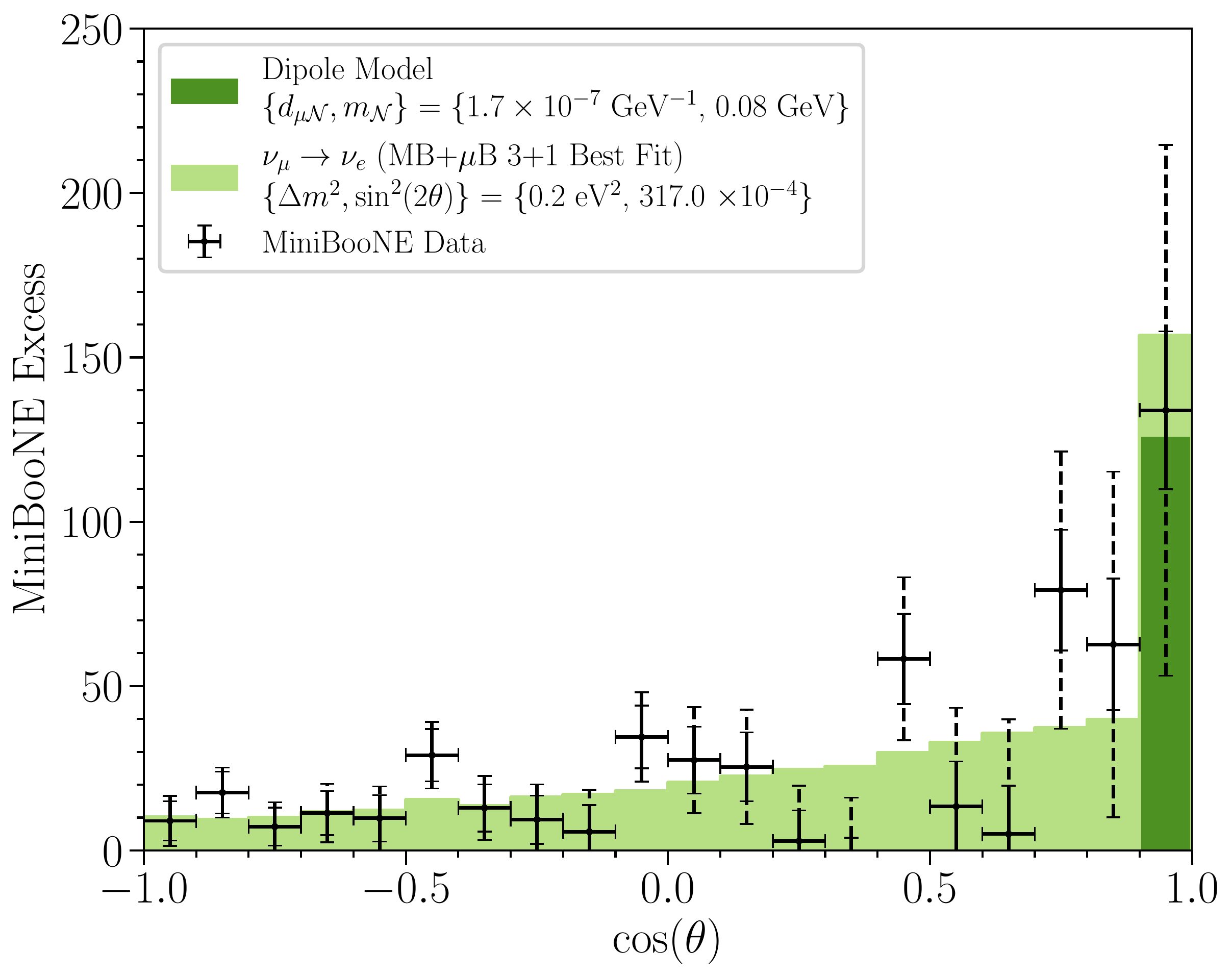}
    \caption{Similar to Fig.~\ref{fig:miniboone_dists}. The $E_\nu^{\rm QE}$ (top) and $\cos \theta$ (bottom) distributions at the example dipole model hypothesis indicated by the black star in Fig.~\ref{fig:miniboone_result_jointosc}. The oscillation component comes from the combined MiniBooNE + MicroBooNE CCQE-like $3+1$ fit~\cite{MiniBooNE:2022emn}.}
    \label{fig:miniboone_dists_jointosc}
\end{figure}

\begin{figure*}[t]
    \centering
    \includegraphics[width=0.45\textwidth]{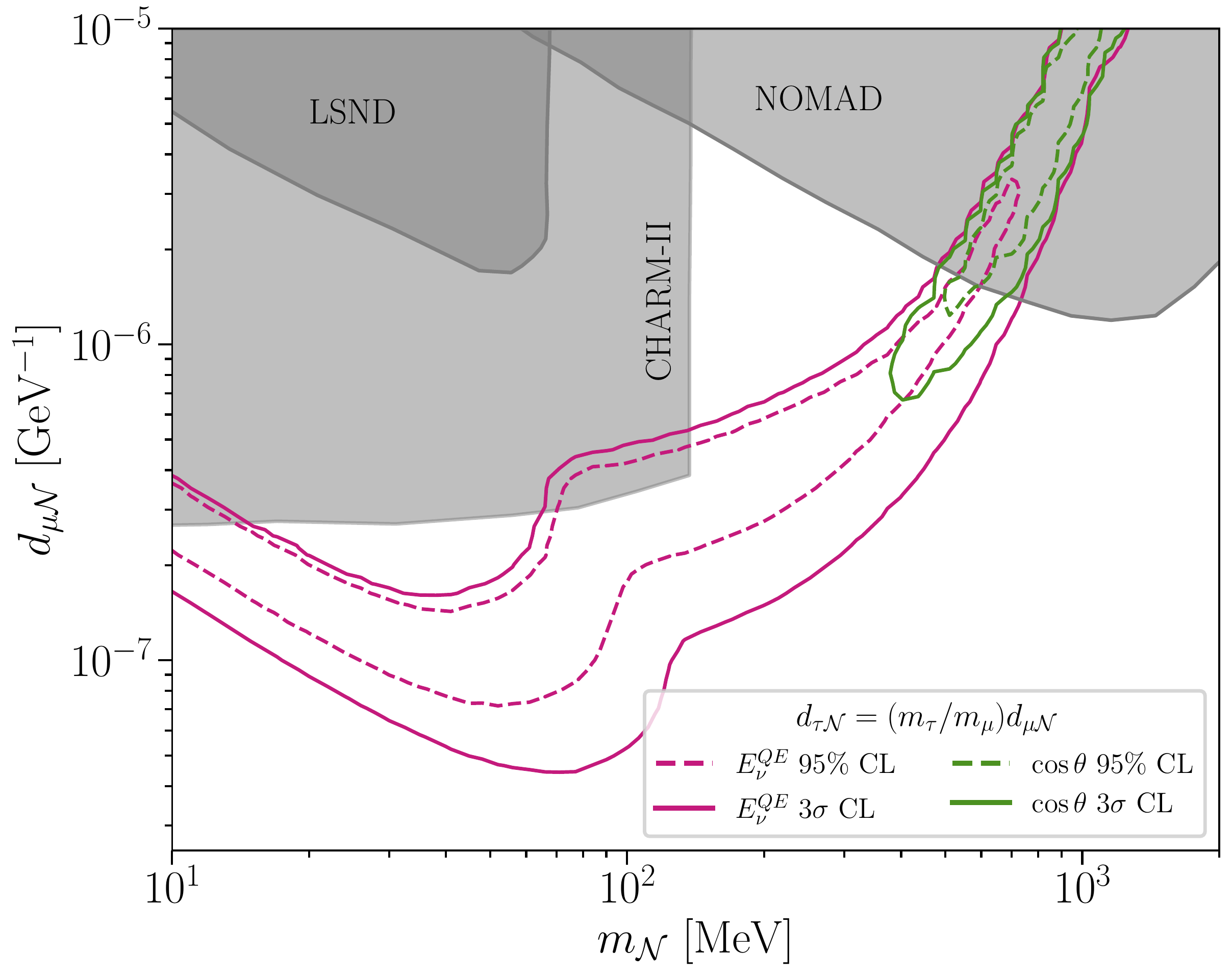}
    \includegraphics[width=0.45\textwidth]{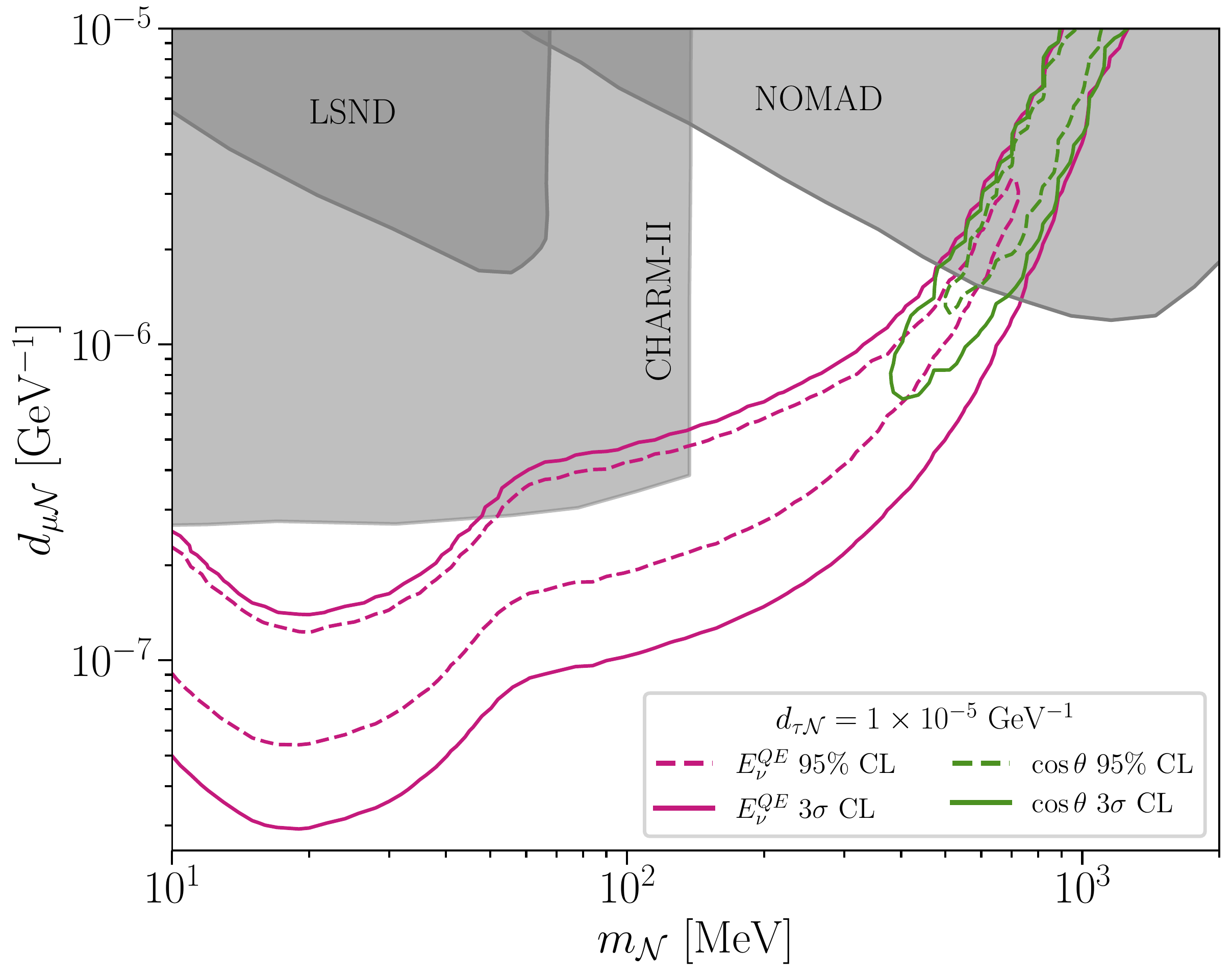}
    \caption{Similar to Fig.~\ref{fig:miniboone_result}, but considering the $\mathcal{N} \to \nu_\tau \gamma$ decay channel. On the left, we show the case with coupling $d_{\tau\mathcal{N}} = (m_\tau/m_\mu) d_{\mu\mathcal{N}}$ and on the right we consider $d_{\tau\mathcal{N}} = 1\times 10^{-5}\;{\rm GeV}^{-1}$ instead. Compared to Fig.~\ref{fig:miniboone_result}, preferred regions in the lower HNL mass region move to lower $d_{\mu\mathcal{N}}$ values. \label{fig:miniboone_result_dtau}}
\end{figure*}

The result from the fit procedure described above is shown in \Cref{fig:miniboone_result}.
One can see that it is difficult to explain the $E_\nu^{\rm QE}$ and $\cos \theta$ distributions through the same dipole-coupled HNL, as the two distributions prefer different regions of dipole parameter space.
The preferred regions overlap at the $2\sigma$ CL, though some of this overlap region is in tension with constraints derived from the NOMAD single-photon analysis~\cite{Gninenko:1998nn}.
As stated above, the difference between this result and the result in Ref.~\cite{Vergani:2021tgc} is driven mainly by the updated form factor.
This reduces the Primakoff upscattering rate at large scattering angles, requiring the fit to push to larger $\mathcal{N}$ masses and dipole couplings to explain this part of the MiniBooNE excess.
That being said, there is a region in parameter space for $d \sim 10^{-6}\;{\rm GeV}^{-1}$ and $m_\mathcal{N} \sim 0.5\;{
\rm GeV}$ which is (1) within the $2\sigma$ allowed region for the $E_\nu^{\rm QE}$ distribution, (2) within the $2\sigma$ allowed region for the $\cos \theta$ distribution, and (3) not ruled out by NOMAD's single-photon search~\cite{Gninenko:1998nn}.
In Fig.~\ref{fig:miniboone_dists} we show $E_\nu^{\rm QE}$ and $\cos \theta$ distributions for an example parameter point in this region, located at $d = 1.25 \times 10^{-6}\;{\rm GeV}^{-1}$ and $m_\mathcal{N} = 0.47\;{\rm GeV}$.
One can see that this model can describe most of the excess except for the region at $\cos \theta \lesssim 0$.

This situation might be improved when considering alternative oscillation scenarios.
The above fits assumed the MiniBooNE-less global-fit result, which found a best-fit solution at $\{\Delta m^2 \approx 1.3\;{\rm eV}^2, \sin^2(2\theta_{\mu e}) \approx 6.9 \times 10^{-4}\}$~\cite{Vergani:2021tgc}.
We now consider an alternative sterile neutrino hypothesis: the result from the recent MiniBooNE + MicroBooNE CCQE-like $3+1$ combined fit performed by the MiniBooNE collaboration~\cite{MiniBooNE:2022emn}.

The MiniBooNE + MicroBooNE CCQE-like $3+1$ combined analysis found a best-fit solution at $\{\Delta m^2 \approx 0.2\;{\rm eV}^2, \sin^2(2\theta_{\mu e}) \approx 0.03\}$~\cite{MiniBooNE:2022emn}.
This introduces a much larger $\nu_\mu \to \nu_e$ oscillation component in MiniBooNE.
Thus, the dipole model is primarily driven to explain the lowest energy and most forward-angle portion of the excess.
A mild preference for a dipole-coupled heavy neutral lepton is found at the $1\sigma$ level.
As the dipole model is no longer required to explain the broad-angle portion of the MiniBooNE excess, the angular fit is able to accommodate a large range of heavy neutrino masses while the energy fit prefers lower heavy neutrino masses at $m_{\mathcal{N}} \lesssim 100\;{\rm MeV}$.
The preferred regions in dipole parameter space under this oscillation hypothesis are shown in Fig.~\ref{fig:miniboone_result_jointosc}.
As a benchmark point, we consider a solution at $d = 1.7 \times 10^{-7}\;{\rm GeV}^{-1}$ and $m_\mathcal{N} = 0.08\;{\rm GeV}$. As shown in Fig.~\ref{fig:miniboone_dists_jointosc}, this benchmark point can reasonably describe the $E_\nu^{\rm QE}$ and $\cos\theta$ distributions of the MiniBooNE excess.


One can also consider a nonzero transition magnetic moment coupling between the $\mathcal{N}$ and the $\nu_\tau$ flavor eigenstate.
This would open up the decay channel $\mathcal{N} \to \nu_\tau \gamma$, increasing the decay width by the ratio $(|d_{\mu\mathcal{N}}|^2 + |d_{\mu\mathcal{N}}|^2)/|d_{\mu\mathcal{N}}|^2$.
This will have a more pronounced impact on the fit in the lower HNL mass region of parameter space, as lifetimes in the higher HNL mass region are sufficiently short such that introducing another decay channel does not appreciably change the phenomenology. 

As discussed in section~\ref{sec:model}, in some UV completions of the dipole model, a natural scaling given by $d_{\tau\mathcal{N}}/d_{\mu\mathcal{N}} = m_\tau/m_\mu$.
The resulting preferred regions in dipole model parameter space under this assumption are shown on the left panel of Fig.~\ref{fig:miniboone_result_dtau}.
One can see that, compared with Fig.~\ref{fig:miniboone_result}, solutions explaining the $E_\nu^{\rm QE}$ distribution have opened at lower $d_{\mu\mathcal{N}}$ couplings for $m_\mathcal{N} \lesssim 100\;{\rm MeV}$.
We also examine the effect of large tau coupling $d_{\tau\mathcal{N}} = 1 \times 10^{-5}\;{\rm GeV}^{-1}$, which is meant to capture the extent of flexibility introduced into the dipole model when allowing for nonzero $d_{\tau\mathcal{N}}$.
The preferred regions for this case are shown in the right panel of Fig.~\ref{fig:miniboone_result_dtau}.
For both cases, we consider an oscillation contribution given by the MiniBooNE-less global fit. 

\section{Neutrissimos at \minerva} \label{sec:mvresults}

Neutrino upscattering can also occur in the \minerva detector. 
We choose to study \minerva for two main reasons:
i) the NuMI beam is a higher-energy beam in comparison with the BNB. 
This is specially true for the medium-energy (ME) NuMI configuration, where $\langle E_\nu \rangle \simeq 7$~GeV, but it is still the case for the low-energy (LE) configuration, where $\langle E_\nu \rangle \simeq 3$~GeV.
This allows us to probe HNLs of larger masses.
ii) it is one of the few accelerator experiments in the few GeV region to have a dedicated neutrino-electron ($\nu-e$) scattering analysis. 
While measurements of this channel have been performed with greater precision at experiments like CHARM~\cite{CHARM:1988tlj} and CHARM-II~\cite{CHARM-II:1994dzw}, LSND~\cite{LSND:2001akn}, reactors~\cite{TEXONO:2009knm}, Borexino~\cite{Bellini:2011rx,Borexino:2017rsf}, and Super-Kamiokande~\cite{Super-Kamiokande:2001ljr}, they are not as well suited for the study of the MiniBooNE explanations considered here, where HNLs have hundreds of MeV in mass. 
With the exception of CHARM and CHARM-II, the previous experiments operate at energies below the HNL production threshold and are therefore not sensitive to our region of interest. 
While we could also consider CHARM and CHARM-II, we note that they observe larger neutrino-induced backgrounds thanks to the faster growth of the SM cross section with respect to the dipole one.
In addition, the HNLs would be produced with a larger boost factor, and therefore escape more often.
\begin{figure}
\includegraphics[width=0.45\textwidth]{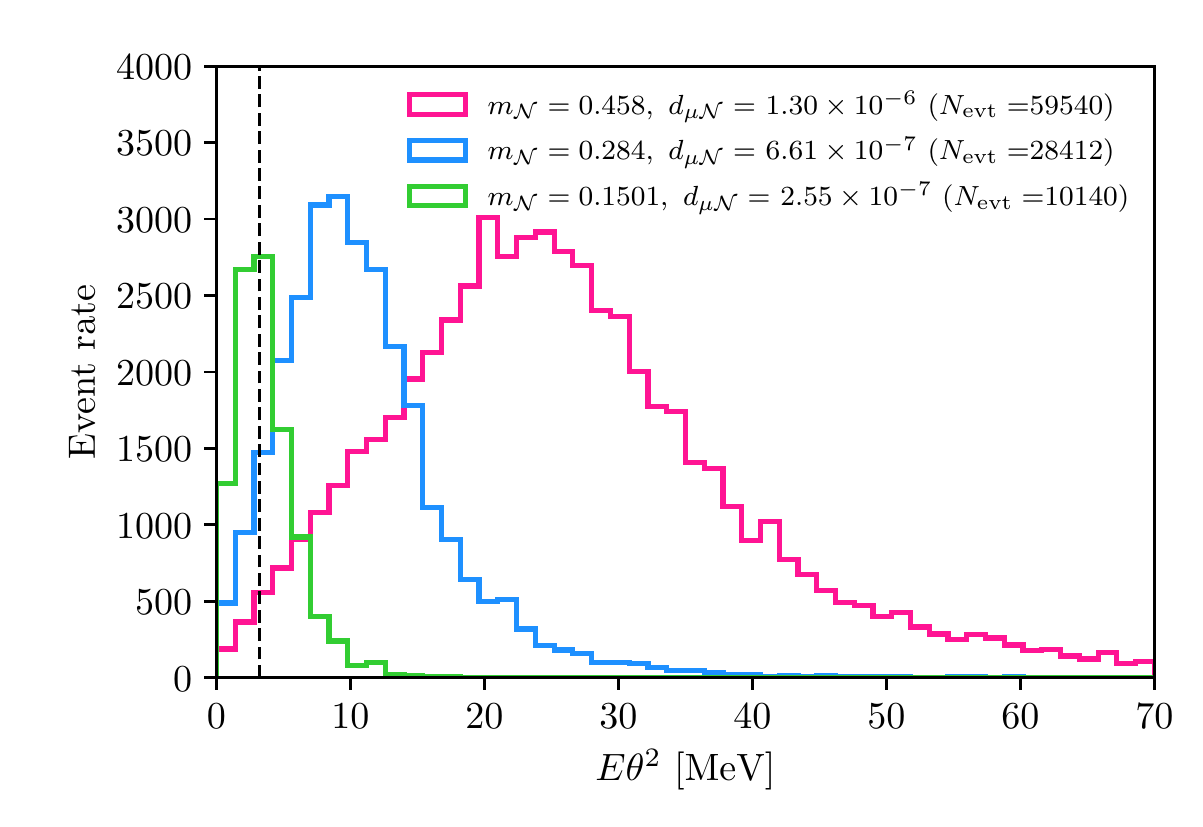}
\includegraphics[width=0.45\textwidth]{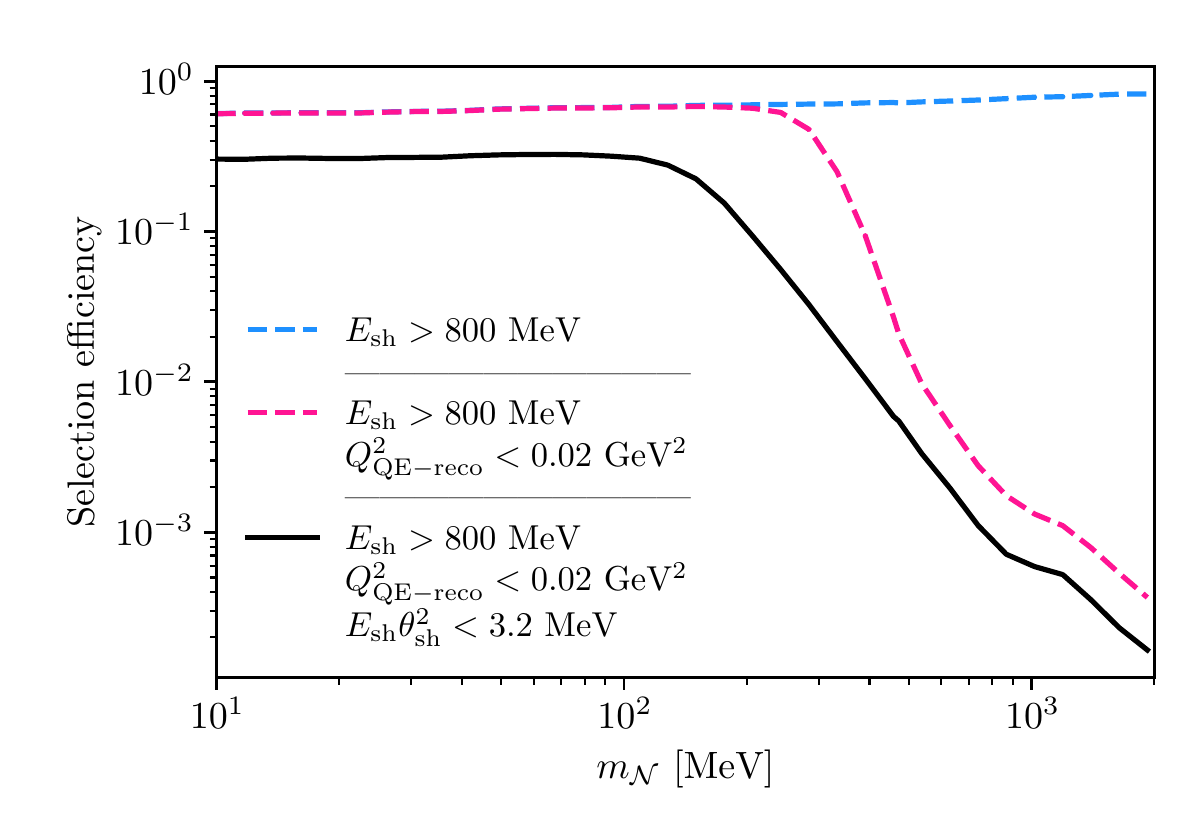}
\caption{
Top) The predicted $E_{\rm sh} \theta_{\rm sh}^2$ distribution before detector smearing and signal selection for three choices of model parameters at \minerva.
Bottom) The signal selection efficiency of our analysis cuts, excluding the $dE/dx$ cut, as a function of the HNL mass.
\label{fig:etheta2}}
\end{figure}

\begin{figure}
\centering
\includegraphics[width=0.45\textwidth]{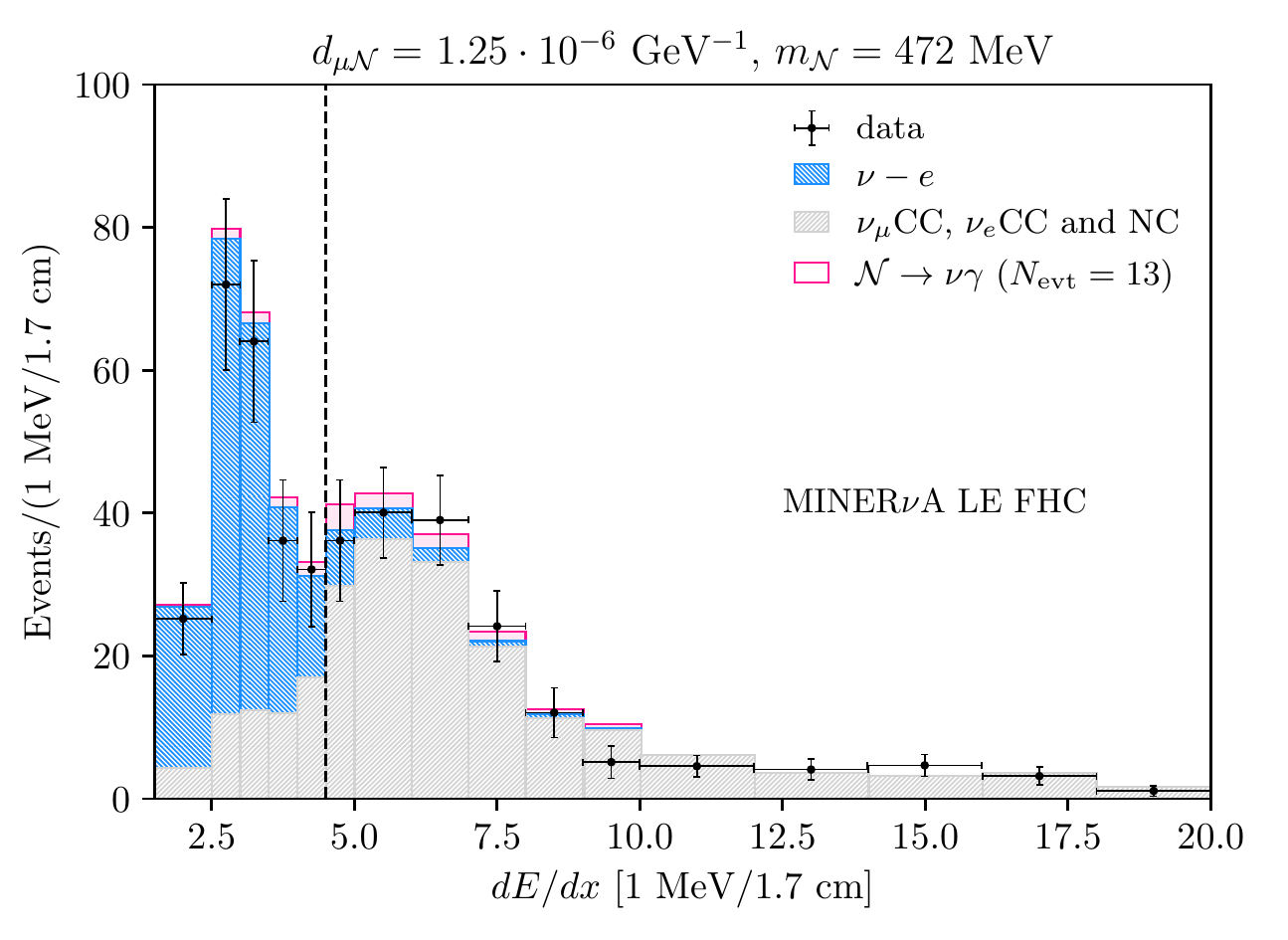}
\includegraphics[width=0.45\textwidth]{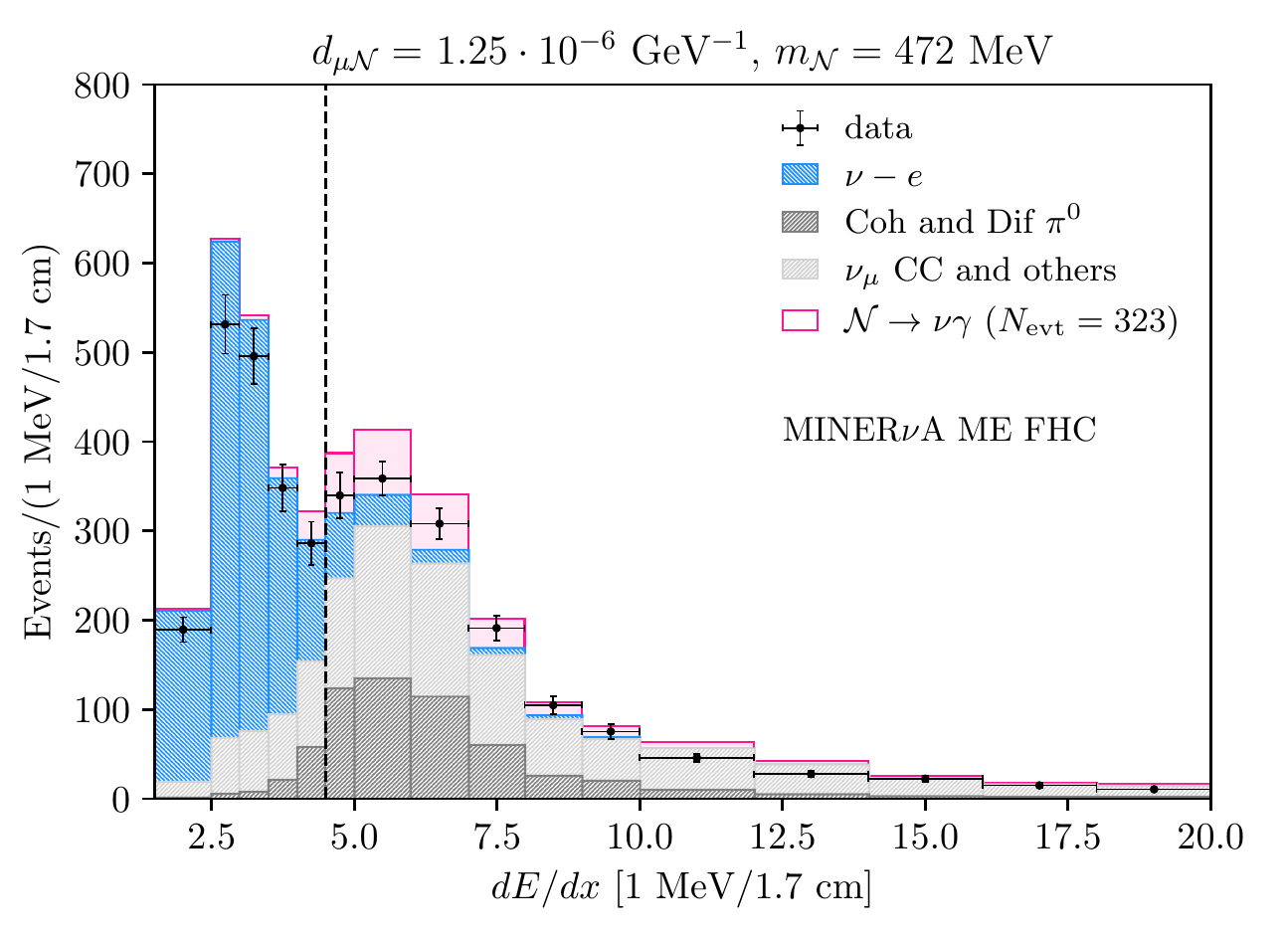}
\includegraphics[width=0.45\textwidth]{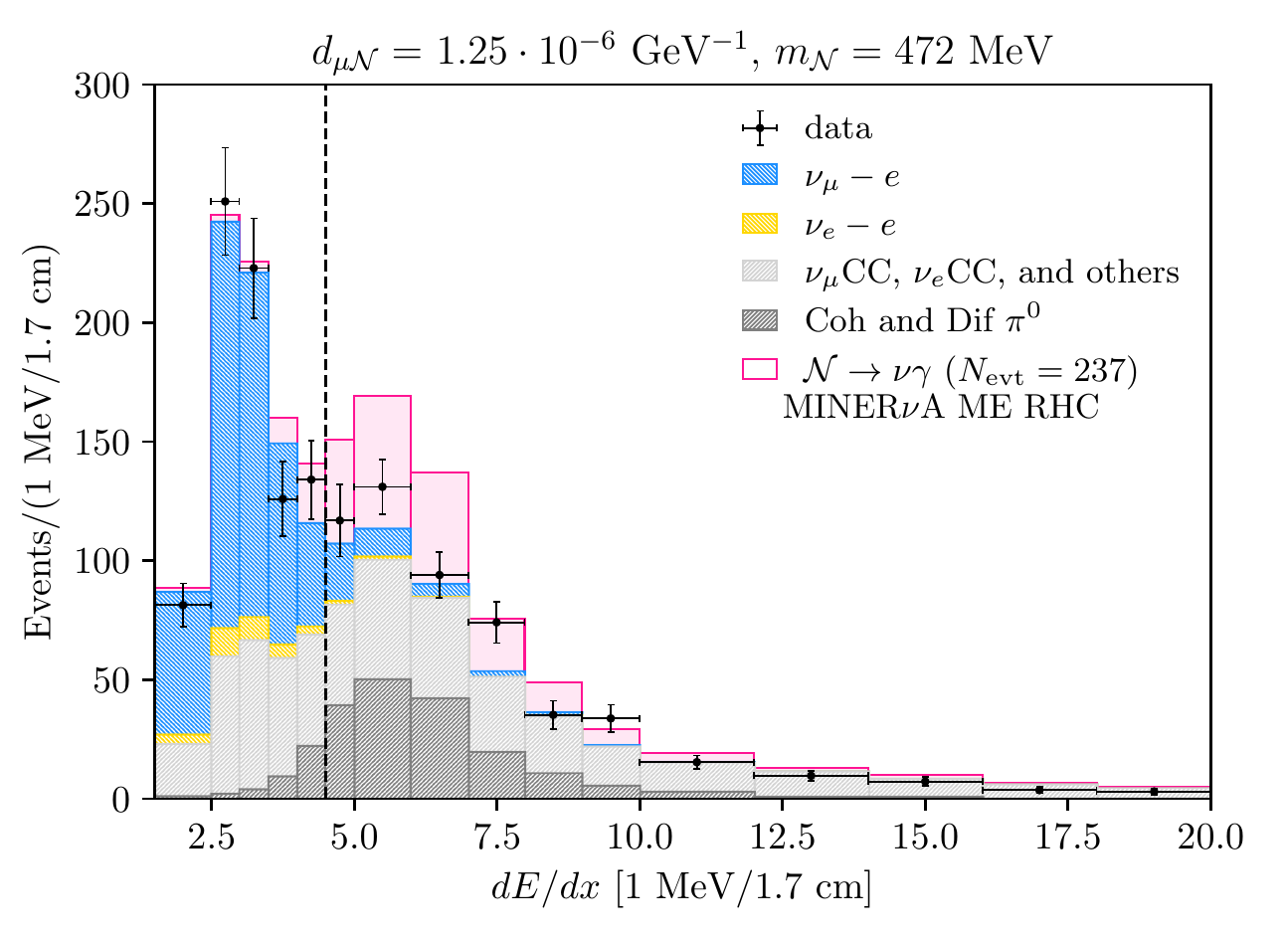}
\caption{The $dE/dX$ distribution of selected events for the three \minerva analyses. From top to bottom: LE FHC, ME FHC, and ME RHC. The latter has the largest sensitivity due to the smaller backgrounds. All distributions are shown post-\minerva tune, except for ME RHC, where it is shown before tuning. \label{fig:dedx}}
\end{figure}

In this work, we will consider three existing measurements of the neutrino-electron cross section by \minerva~\cite{Park:2015eqa,Valencia:2019mkf,MINERvA:2022vmb}. 
The first was performed in the LE configuration of the NuMI beam operating with a forward-horn current (FHC), optimizing the number of neutrinos produced.
The last two were performed in the ME configuration, one in FHC and the other in reverse-horn current (RHC) mode, the latter optimizing the number of antineutrinos.
The ME RHC measurement, also the most recent, is particularly sensitive due to the smaller antineutrino- and neutrino-induced backgrounds.
Unlike the dipole cross section, antineutrino-nucleus weak cross sections are smaller than neutrino-nucleus cross sections.

\subsection{Simulation}\label{sec:simulation_minerva}

\begin{table}[]
    \centering
    \begin{tabular}{|c|c|c|c|}
    \hline
    Target & $z$-location (cm) & $z$-extent (cm) & Mass (kg)  \\
    \hline
    1-Fe & 13.6 & 2.567 & 370  \\  
    1-Pb & 13.6 & 2.578 & 317  \\ 
    \hline
    2-Fe & 31.3 & 2.563 & 370  \\  
    2-Pb & 31.3 & 2.581 & 317  \\ 
    \hline
    3-Fe & 53.4 & 2.573 & 197  \\  
    3-Pb & 53.4 & 2.563 & 141  \\ 
    3-C  & 53.4 & 7.620 & 194 \\ 
    \hline
    Water & 89.5 & 18.06 & 530 \\
    \hline
    4-Pb & 125.6 & 0.795 & 263 \\  
    \hline
    5-Fe & 138.9 & 1.289 & 186 \\
    5-Pb & 138.9 & 1.317 & 162  \\ 
    \hline
    \end{tabular}
    \caption{Specifications of the \minerva nuclear targets as implemented in \texttt{LeptonInjector}~\cite{IceCube:2020tcq} for this analysis. Each nuclear target is defined as a hexagonal prism with an apothem of 92~cm. Z positions and extents of each nuclear target have been taken from Table~4 of Ref.~\cite{MINERvA:2013zvz}. The coordinate system is defined such that $z=0$ corresponds to the front of the \minerva detector. We have confirmed that the fiducial mass (bounded by an 85~cm apothem hexagon) of each nuclear target subcomponent matches the fiducial mass quoted in Table~4 of Ref.~\cite{MINERvA:2013zvz}. The last column of this table refers to the mass of each nuclear target subcomponent within the 92~cm apothem hexagonal prism.}
    \label{tab:detector_components}
\end{table}

\begin{figure*}[t]
\includegraphics[width=0.49\textwidth]{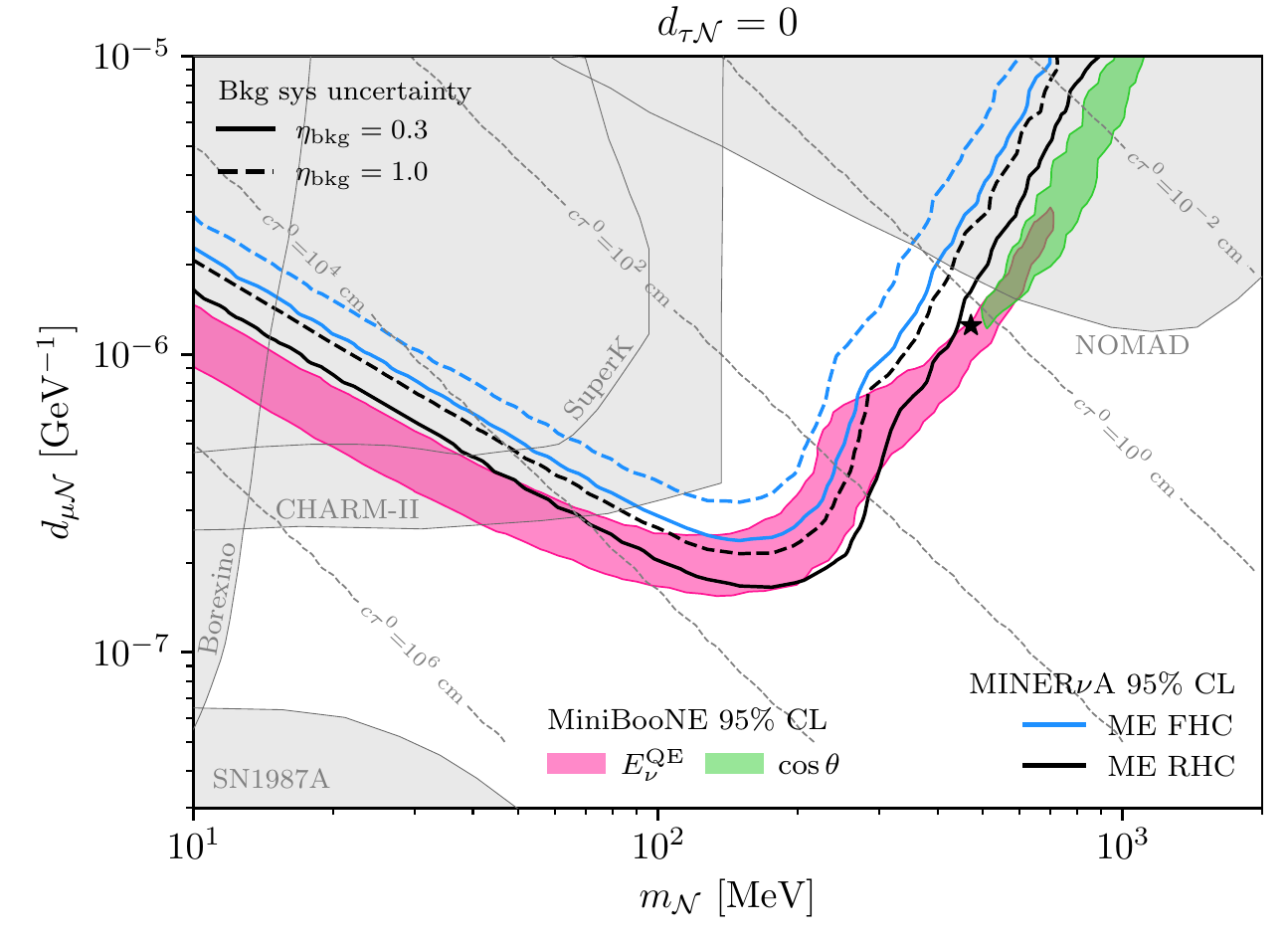}
\includegraphics[width=0.49\textwidth]{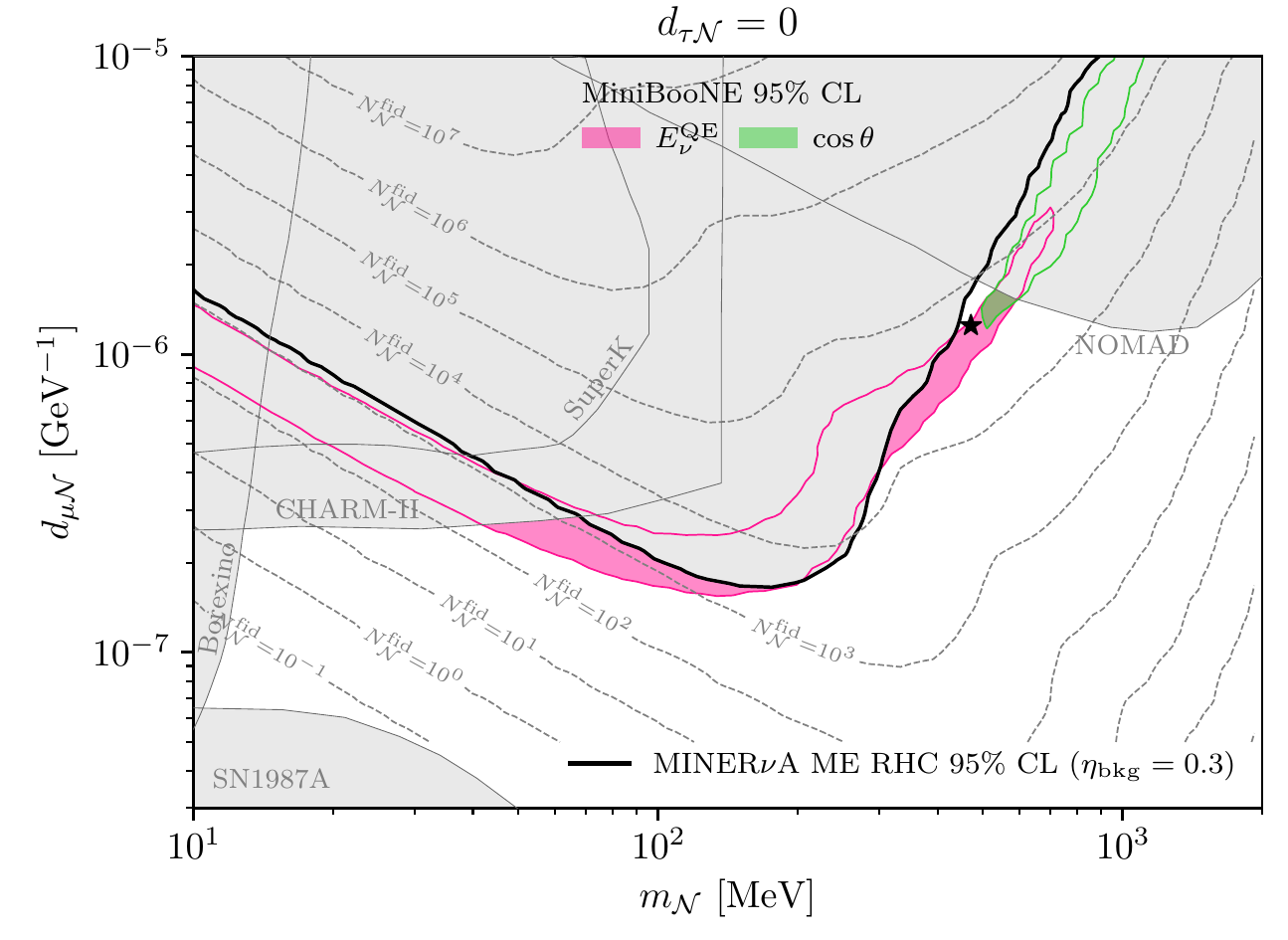}
\caption{Left) \minerva constraints in the dipole parameter space at $95\%$~C.L. Solid lines show our nominal limits assuming a $\eta_{\rm bkg} = 30\%$ Gaussian systematic uncertainty on the background normalization, and dashed ones show the constraints assuming an inflated uncertainty of $\eta_{\rm bkg} = 100\%$ on the background.
Regions of preference to explain MiniBooNE in the minimal dipole model are also shown as filled contours at $95\%$~C.L. 
Right) Contours of constant $N_\mathcal{N}^{\rm fid}$, the total number of new-physics photons that convert inside the fiducial volume, overlaid on top of the same parameter space. \label{fig:final_constraint}}
\end{figure*}

As described in section~\ref{sec:simulation}, we use \texttt{LeptonInjector}~\cite{IceCube:2020tcq} to simulate $\nu A \to \mathcal{N} A$ upscattering inside as well as outside the \minerva detector.
A schematic depiction of this process is shown in the top panel of Fig.~\ref{fig:lepinj_example}.
We include upscattering in the upstream dirt, in the surrounding air, in the nuclear target planes (detailed below), in the plastic scintillator, as well as in the outermost electromagnetic calorimeter.
While most of these components are not part of the fiducial volume for the $\nu-e$ analysis, they can significantly contribute to the signal rate due to the displaced decays of the HNLs.
For long-lived HNLs, upscattering in the dirt dominates the signal rate, followed by the nuclear target planes, which contain high-density materials like $\isotope[208]{Pb}$ and $\isotope[56]{Fe}$.
Detailed modeling of the detector geometry and material composition is necessary to correctly predict the contributions from these different upscattering sites.
We define each nuclear target to be a hexagonal prism with apothem 92~cm, matching that of the tracker region~\cite{Park:2013dax}, and z extent given by Table~4 of Ref.~\cite{MINERvA:2013zvz}.
The six nuclear target planes are detailed in \Cref{tab:detector_components}.
In the bottom panel of Fig.~\ref{fig:lepinj_example}, we show the positional distribution of the upscattering rate within two of the \minerva nuclear targets as simulated using \texttt{LeptonInjector}~\cite{IceCube:2020tcq}.
We also consider upscattering within the electromagnetic calorimeter, defined as a hexagonal prism surrounding the inner detector with 107~cm apothem, and within the steel veto shield $\sim 1$~m in front of \minerva~\cite{Park:2013dax}.
The fiducial volume of \minerva for the $\nu-e$ analyses is assumed to be approximately the same for both the LE and ME analyses and is defined as a hexagon of $81$~cm apothem with $\sim 2.8$~m z extent inside the plastic-scintillator. 

After production, we track the HNL's path through the detector and force a decay to occur before the end of the fiducial volume; we then down-weight the event by the probability of decaying within the considered region.
Each decay produces a photon for which we physically sample a pair-production location.
Events that do not pair-produce within the fiducial volume are removed.
This procedure accounts for events where the HNL decays outside the fiducial volume, but the photon conversion happens inside of it.
This effect is important for short-lived HNLs since the rate of HNLs produced in the high-density lead planes can significantly contribute to the signal rate even though they are not contained in the fiducial volume.

The neutrino fluxes for the LE mode have been taken from \refref{MINERvA:2016iqn} and for the ME they have been digitized from \refref{Bashyal:2021tzd}. 
The total exposures used in the three $\nu-e$ analyses are $3.43 \times 10^{20}$~POT for LE-FHC, $1.16\times10^{21}$~POT for ME-FHC, and $1.22 \times 10^{21}$~POT for ME-RHC.
\begin{figure*}[t]
\includegraphics[width=0.49\textwidth]{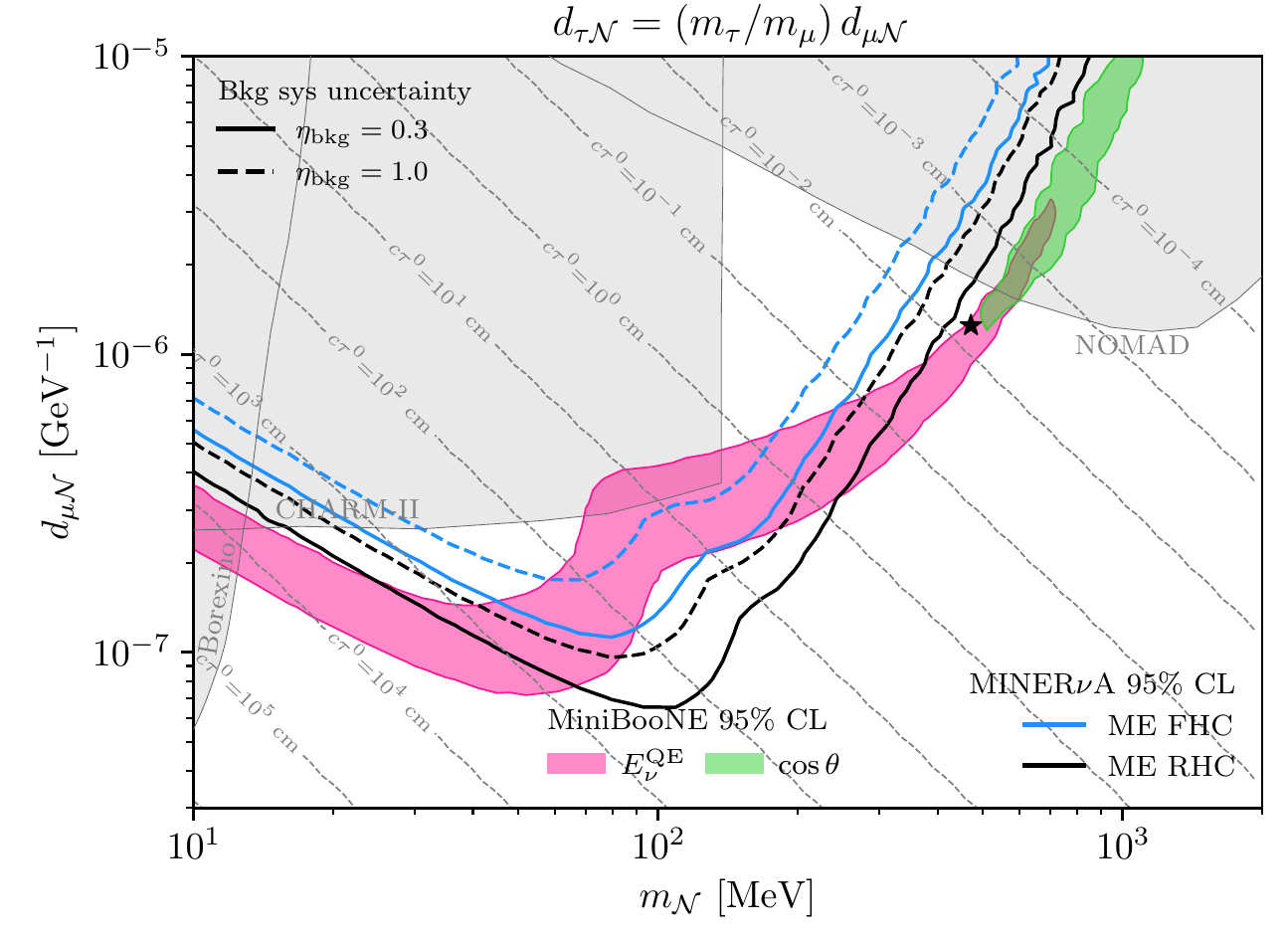}
\includegraphics[width=0.49\textwidth]{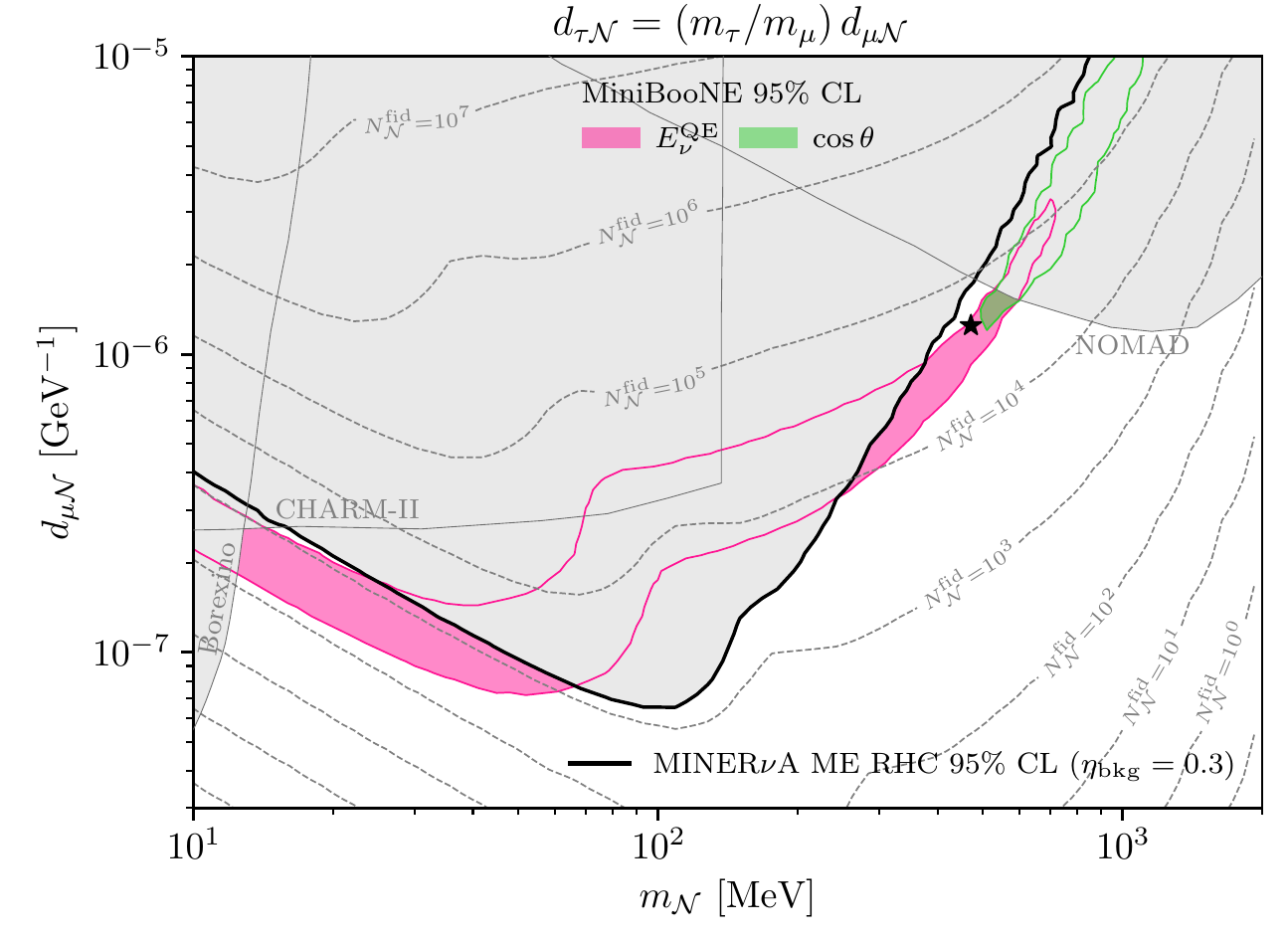}
\caption{Same as \Cref{fig:final_constraint} but for the case where HNLs have a larger tau-neutrino dipole following an approximate scaling of UV completions of the operator in \Cref{eq:dim6_dipoles}, $d_{\tau \mathcal{N}} = m_\tau/m_\mu \times d_{\mu \mathcal{N}}$.
The HNL is shorter-lived due to the additional decay $\mathcal{N}\to\nu_\tau \gamma$.\label{fig:constraint_2}}
\end{figure*}

To reduce neutrino-induced backgrounds, \minerva applies an extensive list of selection cuts.
To properly estimate the resulting efficiency of these cuts in our HNL signal, it is important to correctly model the reconstruction of the energy and angle of the single photons.
In the absence of a full detector simulation, we proceed to approximate the detector energy and angular resolutions as Gaussian functions.
For the energy resolution, we take $\sigma_{E}/E  = 5.9\% /(\sqrt{E/{\rm GeV}}) + 3.4\%$~\cite{Park:2015eqa}, while for the angular resolution, we take an energy-independent angular resolution of $\sigma_\theta = 0.7 ^\circ$, assumed to be isotropic in the shower's azimuthal angle~\footnote{This is only an approximation, as the \minerva detector is not azimuthally symmetric. Nevertheless, the differences in resolution in the X and Y planes are small~\cite{Park:2013dax,Valencia-Rodriguez:2016vkf} and neglected here.}.
The angular resolution is implemented by sampling a polar angle $\delta \theta$ from a Gaussian distribution of standard deviation $\sigma_\theta$, rotating the photon by $\delta \theta$ with respect to its momentum, assigning it an azimuthal angle $\phi$ from the uniform distribution $[0,2\pi]$, and finally rotating the photon back to the laboratory frame by its original polar angle $\theta_{\rm true}$.

\subsection{Event selection}\label{sec:selection_minerva}

Now we discuss the most important signal selection cuts. 
The analyses~\cite{Park:2015eqa,Valencia:2019mkf} make use of a long list of signal selection cuts, designed to suppress as many neutrino-nucleus scattering backgrounds as possible.
The most worrisome backgrounds include $\pi^0$ production and $\nu_e$CC scattering. 
The former is particularly important for our radiative decay signal, as it can give rise to coherent single-photon-like signatures. 
Cuts related to the shower radius and transversal as well as longitudinal profiles are not implemented in our analysis but are expected to have large acceptance due to our signal being a true single photon (as opposed to two photons from $\pi^0$ or from the $e^+e^-$ pairs considered in the new physics model of \refref{Arguelles:2018mtc}).
The series of cuts are illustrated in \Cref{fig:etheta2}, where we show the acceptance of the cuts as a function of the HNL mass.
The selection acceptance is largely independent of the dipole coupling.

We start with the cut on the reconstructed shower energy, $E_{\rm sh} > 0.8$~GeV.
To suppress $\nu_e$CC backgrounds, a cut on the reconstructed momentum exchange under the hypothesis of neutrino-nucleon quasi-elastic scattering is also implemented. 
It is defined a s $Q^2_{\rm QE-reco} = 2 m_n (E_\nu^{\rm QE-reco} - E_{\rm sh})$, with 
\begin{equation}
E_\nu^{\rm QE-reco} = \frac{m_n E_{\rm sh} - m_e^2/2}{m_n - E_{\rm sh} + p_e \cos{\theta_e}},
\end{equation}
and the analysis requires $Q^2_{\rm QE-reco} < 0.02$~GeV$^2$. 
We note that for $\nu-e$ scattering this cut can be understood by the following relation,
\begin{equation}
    Q^2_{\rm QE-reco} \simeq (E_{\rm sh} \theta_{\rm sh})^2  < \frac{m_e E_\nu}{2},
\end{equation}
where we dropped higher-order terms in electron mass and $\theta_{\rm sh}$.
For neutrino-nucleus upscattering in the forward direction (small $\theta_\mathcal{N}$), two-body kinematics for an infinitely heavy nucleus gives $Q^2_{\rm QE-reco} \simeq (E_\mathcal{N} \theta_\mathcal{N})^2 < (E_\nu m_A -  m_{\mathcal{N}}^2)/2$, which is a much looser constraint. In addition, the decay of $\mathcal{N}$ introduces even more spread in the angular distribution, so we can already expect the cut on $Q^2_{\rm QE-reco}$ to be very important.

\begin{figure*}[t]
\includegraphics[width=0.49\textwidth]{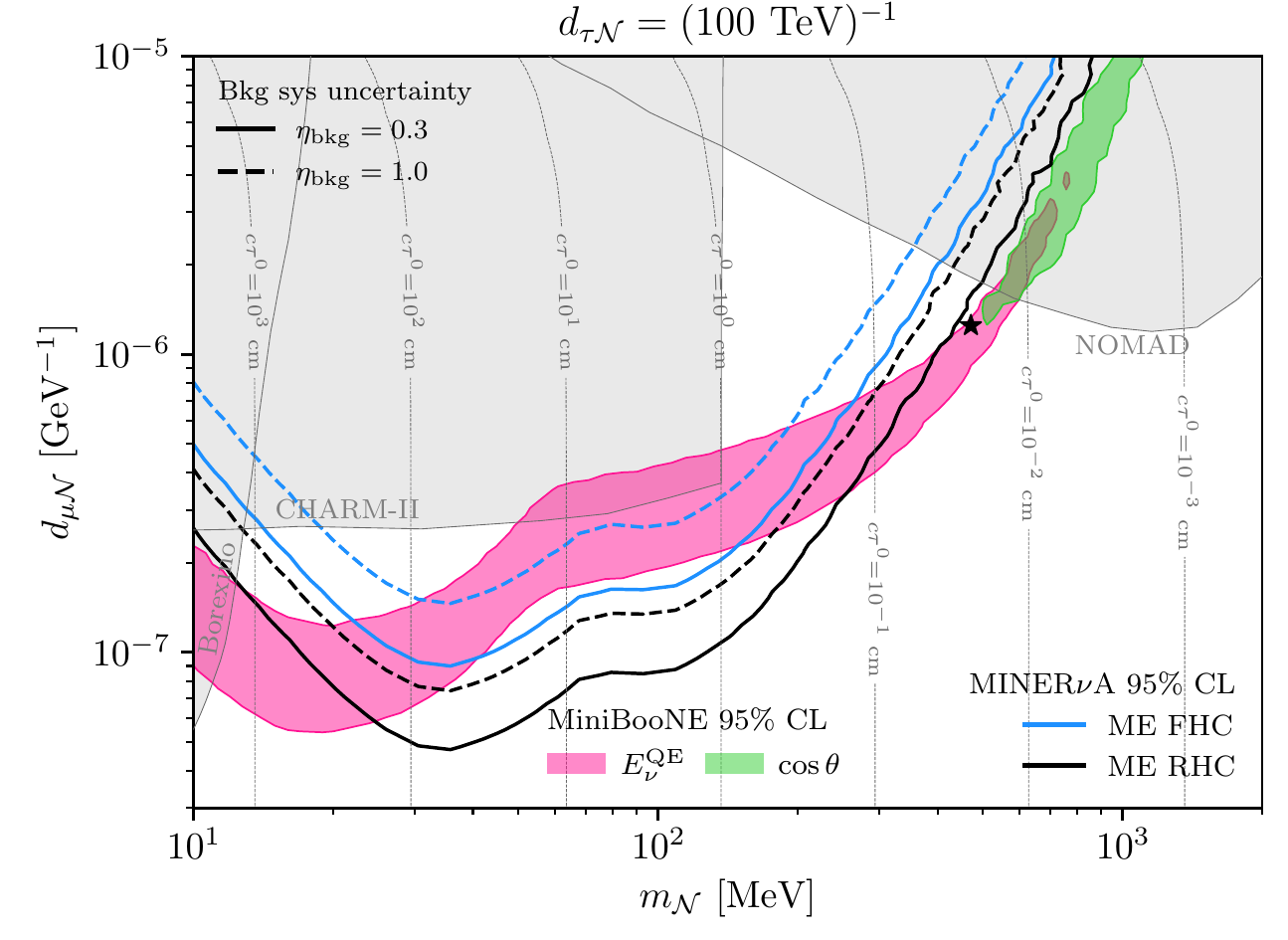}
\includegraphics[width=0.49\textwidth]{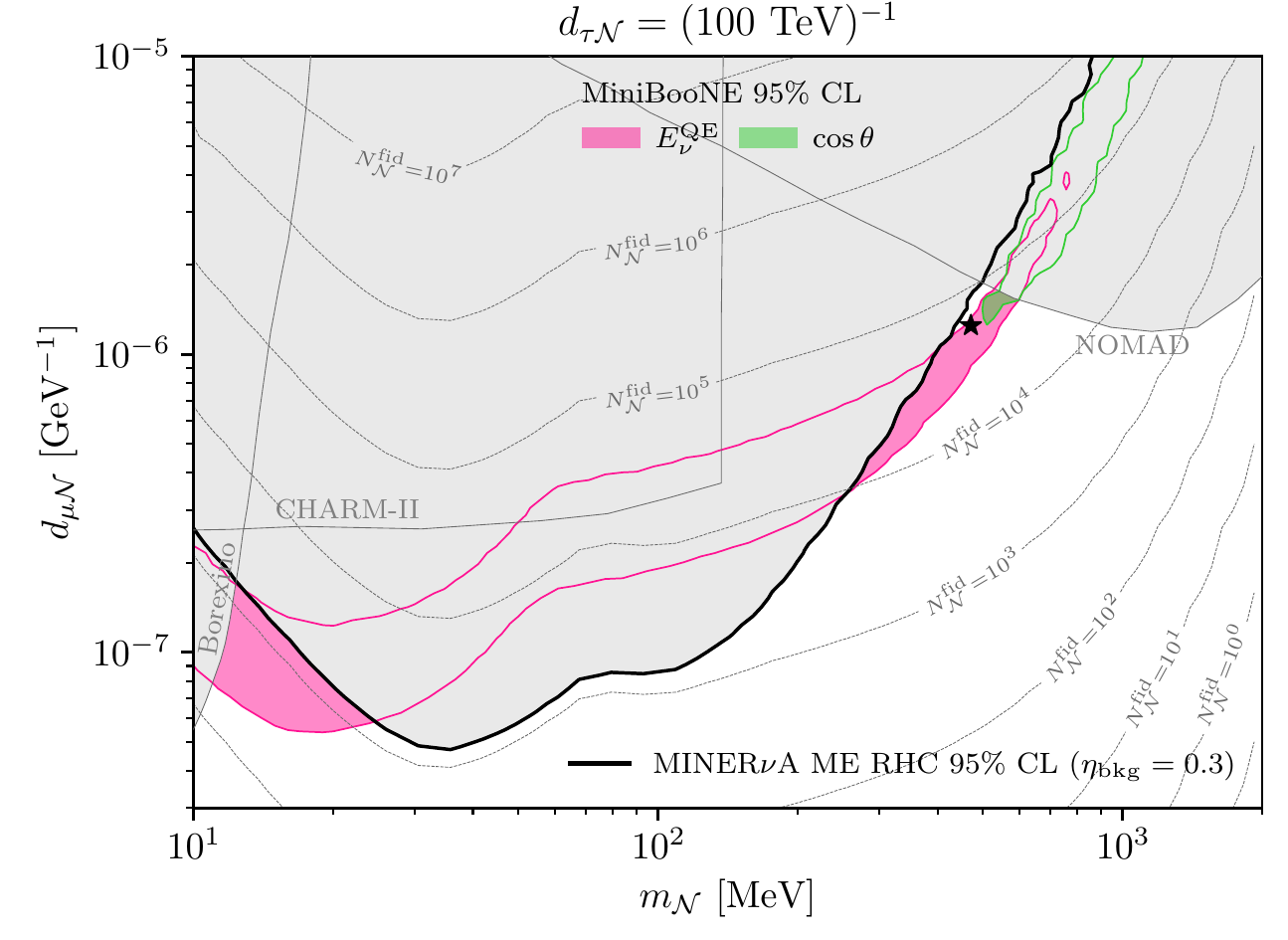}
\caption{Same as \Cref{fig:final_constraint} but for the case where HNLs have a larger and fixed tau-neutrino dipole, $d_{\tau \mathcal{N}} = (100 \text{ TeV})^{-1}$.
The HNL is shorter-lived due to the additional decay $\mathcal{N} \to\nu_\tau \gamma$.\label{fig:constraint_3}}
\end{figure*}

The most stringent cut in the analysis; however, is in $E_{\rm sh} \theta_{\rm sh}^2$, required it to be $< 3.2$~MeV rad$^2$. 
The acceptance of this selection varies from under $10^{-4}$ at high masses to approximately $30\%$ at the lowest masses. 
It is larger for Majorana than Dirac HNLs, as Dirac HNLs are more likely to produce backward-going photons.
Finally, the last cut we implement is the cut on the mean $dE/dx$ of first four scintillator planes, requiring $dE/dx > 4.5$~MeV$/1.7$~cm. 
\Cref{fig:dedx} shows the $dE/dx$ distributions of the SM background and HNL decay events, as well as a dashed line representing the $dE/dx$ cut.
The acceptance of this cut is the least-understood aspect of our analysis as we do not have access to a full detector simulation.
The shape of our $dE/dx > 4.5$~MeV$/1.7$~cm distribution is assumed to be identical to that of the coherent and diffractive $\pi^0$ backgrounds shown in \Cref{fig:dedx}.
In this approximation, we find this cut has an acceptance of $93\%$ and $91\%$ for FHC and RHC modes, respectively. 

\subsection{Results}\label{sec:results_minerva}

The final $95\%$~C.L. constraints on the dipole model are shown in \Cref{fig:final_constraint}.
The MiniBooNE regions of preference are also shown for comparison.
At the lowest values of $m_{\mathcal{N}}$ where the HNLs are long lived, our constraints are less sensitive to the MiniBooNE best-fit region than at higher masses due to the larger HNL boost factors at \minerva as well as the smaller fiducial volume when compared to MiniBooNE.
For lifetimes longer than $c\tau^0\sim 100$~cm, the event rate in both experiments is dominated by dirt upscattering.
For shorter lifetimes, the event rate is dominated by upscattering within the detector itself.

One can see that the constraints from \minerva begin to rule out disfavor regions of parameter space preferred by MiniBooNE.
However, the strongest \minerva $2\sigma$ CL limits presented, which come from the ME RHC measurement and assume 30\% uncertainty on the background normalization, do not rule out the intersection of the $2\sigma$ CL preferred regions from the MiniBooNE $E_\nu^{\rm QE}$ and $\cos\theta$ distributions.
This is because the $\mathcal{N} \to \nu \gamma$ acceptance in the \minerva ES analysis decreases rapidly for larger HNL masses, as shown in \Cref{fig:etheta2}.
In the right panel of \Cref{fig:final_constraint}, we show contours of constant event rate from dipole-coupled HNL decays in \minerva.
A dedicated single-shower analysis improving the acceptance for larger HNL masses would likely be sensitive to the entire region of parameter space preferred by MiniBooNE.

In the left panel of \Cref{fig:final_constraint}, we also show conservative constraints on this model assuming 100\% uncertainty on the background normalization.
This is meant to address the large scale factors (up to factors of $\sim 2$) which have been applied to the high $dE/dx$ backgrounds in the official \minerva analysis~\cite{MINERvA:2022vmb}.
These scale factors come from a tuning procedure in kinematic sideband regions, a process that could potentially wash out any signal from neutrissimo decays.
An optimal analysis would perform a joint fit to both neutrissimo decays and SM high $dE/dx$ backgrounds to derive constraints (and potentially allowed regions) on the neutrissimo model presented here; however, such an analysis is out of the scope of this paper.
We also note that the MINERvA analysis does not include single-photon backgrounds such as radiative $\Delta(1232)$ decays and coherent single photons. 
These components are expected to be small in the energy region of $E_{\rm sh} > 800$~MeV~\cite{Rein:1981ys,Wang:2014nat}, but their inclusion can only make our limits stronger.

We also point out \refref{MINERvA:2016uck}, in which the \minerva collaboration investigated an excess in the high $dE/dx$ sideband region of a $\nu_e$ charged-current quasielastic scattering sample.
Using topological variables related to the shower structure, \minerva concluded that the excess looked more like coherent or diffractive $\pi^0$ production than single photons.
This might suggest that the scale factors in \refref{MINERvA:2022vmb} could also be attributed to additional $\pi^0$ events.
However, the analysis presented here suggests that \minerva may have unique sensitivity to a neutrissimo-based explanation of the MiniBooNE excess, thus motivating a more careful separation of one and two photon events in the high $dE/dx$ region of the \minerva elastic scattering samples.

We also derive constraints considering nonzero $d_{\tau\mathcal{N}}$ in \Cref{fig:constraint_2} and \Cref{fig:constraint_3}.
As expected, this impacts the constraints most significantly at lower HNL masses.
For both $d_{\tau\mathcal{N}} = (m_\tau/m_\mu)d_{\mu\mathcal{N}}$ and $d_{\tau\mathcal{N}} = (100\;{\rm TeV})^{-1}$ The \minerva constraints rule out a large chunk in the middle of the region preferred by the MiniBooNE $E_\nu^{\rm QE}$ distribution.
The constraints do not change for $m_\mathcal{N} \gtrsim 200\;{\rm MeV}$, thus the $2\sigma$ overlap between between the MiniBooNE $E_\nu^{\rm QE}$ and $\cos\theta$ distributions remains valid.

\section{Discussion} \label{sec:discussion}

A number of other existing and planned neutrino experiments are sensitive to an MeV-scale dipole-coupled HNL. 
Super-Kamiokande can look for single photon decays from atmospheric neutrinos which upscatter into HNLs within the Earth~\cite{Gustafson:2022rsz}.
Similarly, one can look for single-photon decays from neutrinos that upscatter into HNLs within the Earth and propagate to a large-scale terrestrial detector such as Super-Kamiokande or Borexino~\cite{Plestid:2020vqf}.
Constraints from these searches are sensitive to longer-lived HNLs, with typical masses $\lesssim 10\;{\rm MeV}$ ($\lesssim 100\;{\rm MeV}$) in the solar (atmospheric) case. 
These constraints no longer apply for the two cases of nonzero $d_{\tau\mathcal{N}}$ which we consider in this work, as the HNL lifetime will be too short to reach the detector.
Observed neutrino interactions from Supernova 1987A can also be used to derive constraints on the dipole model, as significant upscattering would enhance the stellar cooling rate, decreasing the neutrino flux observed on Earth~\cite{Magill:2018jla}.
These constraints require the HNL to be sufficiently long-lived that it can escape the stellar environment; thus, bounds from Supernova 1987A also do not apply when we consider nonzero $d_{\tau\mathcal{N}}$.

We also show constraints derived from the NOMAD search for neutrino-induced single photons, recast as bounds in this parameter space in \refref{Gninenko:1998nn}. 
We rescale them, however, by an overall factor of $\sqrt{2}$ so as to reflect the decay rate we derived in~\Cref{eq:dipole_decay}.
In addition, we note that these limits have been obtained with a much less sophisticated simulation than the ones performed here and that the signal was derived using only events in the pre-shower detector of NOMAD.
In Ref.~\cite{Gninenko:1998nn}, it is suggested that stronger limits could be obtained by considering upscattering locations beyond the pre-shower detector and extended detector volumes where the HNL decay could take place.

We also include the limit imposed by CHARM-II as derived in \refref{Coloma:2017ppo}; however, we note that it was obtained with a simplified procedure. 
The experimental precision on the total neutrino-electron scattering cross section was used to set limits on the total neutrino-electron upscattering ($\nu e \to \mathcal{N} e$) cross sections.
This is a reasonable assumption at low values of $m_{\mathcal{N}}$, but potentially breaks down at values close to the threshold due to differences in kinematics.
A similar constraint can be set using the LSND elastic scattering measurement~\cite{Magill:2018jla}, though it is, in general, less sensitive than the CHARM-II measurement.
A robust re-evaluation of dipole model constraints from these electron scattering measurements is out of the scope of this paper.
In addition to scattering on electrons, CHARM-II can provide new limits in the region of interest by considering coherent neutrino-nucleus upscattering followed by HNL decays into single photons~\cite{Arguelles:2018mtc}. 
The sideband with large $dE/dX$ and large values of $E\theta^2$ can be used to set limits, as proposed in \cite{Arguelles:2018mtc}, although we do not expect them to be as sensitive due to larger backgrounds and larger boosts.

One can also derive constraints in the dipole-coupled HNLs from LEP through the $e^+ e^- \to \mathcal{N} \nu_\ell$, which can proceed through either the $\gamma$ or $Z$ mediators~\cite{Delgado:2022fea}.
However, these constraints require a strong enhancement of the mixing between the HNL and SM neutrino; as we consider such a mixing to be negligible in this model, we do note include constraints from LEP in our results.

We now discuss the potential for future constraints on dipole-coupled HNLs from planned measurements.
Just like the \minerva constraint derived in this work, a neutrino elastic scattering measurement from the NO$\nu$A experiment would be sensitive to the dipole model~\cite{Bian:2017axs}.
This is especially true of the DUNE experiment, which has the potential to make a high-statistics neutrino-electron scattering measurement~\cite{Marshall:2019vdy}.
This would be particularly advantageous for the THEIA@DUNE configuration~\cite{Theia:2019non} due to its low threshold and large volume.
Dedicated searches at neutrino experiments can further improve sensitivity to this model.
As discussed above, a \minerva single-shower analysis without a stringent $E\theta^2$ cut could set much stronger constraints.
Experiments which measure CE$\nu$NS, such as COHERENT, NUCLEUS, and Coherent CAPTAIN-Mills, would also be sensitive to the dipole model by looking for the coincidence of nuclear recoil from Primakoff upscattering and a single photon from the HNL decay~\cite{Bolton:2021pey}.
These experiments would be most sensitive to lower mass HNLs with $m_\mathcal{N} \lesssim 10\;{\rm MeV}$ due to the lower energy of typical neutrino sources for CE$\nu$NS experiments.
Existing and upcoming short baseline neutrino experiments, including MicroBooNE and SBND, have the potential to be sensitive to the neutrissimo model presented in this paper through a dedicated search for single photon events~\cite{Magill:2018jla}.
Additionally, neutrino telescopes like IceCube and KM3NeT could perform searches for events with a double-bang topology from the upscattering and decay of the HNL, reaching sensitivities of $d_{\mu \mathcal{N}} \sim 10^{-7}\;{\rm GeV}^{-1}$ for $m_\mathcal{N} \lesssim 1\;{\rm GeV}$~\cite{Coloma:2017ppo}.
High-energy astrophysical tau neutrino observatories such as TAMBO might also be sensitive to HNL decays from $\nu_\tau$ upscattering~\cite{Romero-Wolf:2020pzh}.
Projections for DUNE~\cite{Atkinson:2021rnp,Schwetz:2020xra} estimate that, in the absence of backgrounds, a search for events with a double-bang morphology could reach $d_{\mu \mathcal{N}}$ values as low as $\mathcal{O}(10^{-8} \text{ GeV}^{-1})$.
Finally, nuclear emulsion and liquid argon detectors at a future LHC Forward Physics Facility (FPF) will also be sensitive to transition magnetic moments between HNLs and SM neutrinos~\cite{Ismail:2021dyp}.

\section{Conclusion} \label{sec:conclusion}
In this work, we have explored a mixed model comprising an eV-scale sterile neutrino and an MeV-scale dipole-coupled HNL.
The former facilities oscillations at short baselines, while the latter introduces the interactions shown in Fig.~\ref{fig:feynman}. 
The dipole-coupled HNL provides an alternative explanation of the MiniBooNE excess to the eV-scale sterile neutrino. 
Thus one can remove MiniBooNE from global 3+1 fits, reducing tension between appearance and disappearance experiments while retaining an explanation of the LSND anomaly~\cite{Vergani:2021tgc}. 
We take the result of the MiniBooNE-less 3+1 global fit as the oscillation contribution to the MiniBooNE excess and attribute the remaining excess to decays of the dipole-coupled HNL.
We find that spectral fits to the $E_\nu^{\rm QE}$ and $\cos \theta$ distributions prefer different regions of parameter space in general, though solutions exist which are compatible with both distributions at the $2\sigma$ confidence level. 

We have also derived constraints on the dipole-coupled HNL model using a \minerva neutrino-electron elastic scattering measurements~\cite{Park:2013dax,Valencia:2019mkf,MINERvA:2022vmb}. 
We find that the most sensitive $\nu-e$ scattering constraints are those obtained with the NuMI medium-energy mode in antineutrino-enhanced beam configuration.
The constraints from antineutrino-mode are especially strong due to a reduction in backgrounds at high $dE/dx$, where we expect HNL decays to contribute. 
As shown in Fig.~\ref{fig:final_constraint}, \minerva can exclude large regions of parameter space preferred by MiniBooNE, but it does not fully exclude it.
There are still allowed MiniBooNE regions at the $2\sigma$ confidence level.
The first is at small $m_{\mathcal{N}}$ values, where HNLs are long-lived and \minerva's small fiducial volume and larger energies reduce the sensitivity.
The second is at larger HNL masses, where the stringent $E\theta^2$ cuts reject most new physics events where decay photons tend to have larger $E\theta^2$. 
We note that a dedicated search at \minerva using the same fiducial volume could significantly improve the signal efficiency in this large-mass region, and would likely have much better sensitivity, and potentially probe the entire MiniBooNE-preferred region, as shown in the right panel of Fig.~\ref{fig:final_constraint}. 
Nevertheless, as it stands, this mixed model of oscillations and decay \href{https://www.youtube.com/watch?v=uBxMPqxJGqI}{is not dead yet.}

\section*{Acknowledgements}

MHS is supported by NSF grant PHY-1707971. 
NSF grant PHY-1801996 supported 
CAA, JMC, AD, and NWK for this work.
Additionally, CAA is supported by the Faculty of Arts and Sciences of Harvard University and the Alfred P. Sloan Foundation.
NWK is supported by the NSF Graduate Research Fellowship under Grant No. 1745302.
MAU is supported by the Department of Physics at the University of Cambridge and SV is supported by the STFC. 
MH was supported by Perimeter Institute for Theoretical Physics. Research at Perimeter Institute is supported by the Government of Canada through the Department of Innovation, Science and Economic Development and by the Province of Ontario through the Ministry of Research, Innovation, and Science. 

\appendix

\section{Upscattering cross section}
\label{app:xsecs}

The cross section for $\mathcal{N}$ production by neutrino upscattering, $\nu_\alpha + A \to \mathcal{N} + A$, has been computed several times in the literature~\cite{Vogel:1989iv,Harnik:2012ni,Brdar:2020quo}. We note, however, that all expressions we could find do not take into account the polarization of the outgoing HNL. This effect is only important when $m_\mathcal{N}/E$ becomes appreciably large and we find it to be a marginal effect in our calculations.

In the massless limit, a beam of left-handed polarized neutrinos will always upscatter to right-handed polarized HNLs, assuming the process takes place purely via the transition magnetic moment. This follows from the chiral structure of the vertex, $\overline{\nu_L} \sigma^{\mu\nu} \mathcal{N}_R$. However, in the massive case, spin and helicity are not equivalent, and both helicity states of $\mathcal{N}$ can be produced. The helicity-flipping channel, $\nu_{h} \to \mathcal{N}_{-h}$, typically dominates, while the helicity-conserving case, $\nu_{h} \to \mathcal{N}_{h}$, will be suppressed by powers of $m_\mathcal{N}/E$, vanishing in the massless limit. Here $h=\pm 1$ denotes the particle's helicity and $E$ is the typical energy scale of the scattering process.

\begin{figure*}
    \centering
    \includegraphics[width=0.49\textwidth]{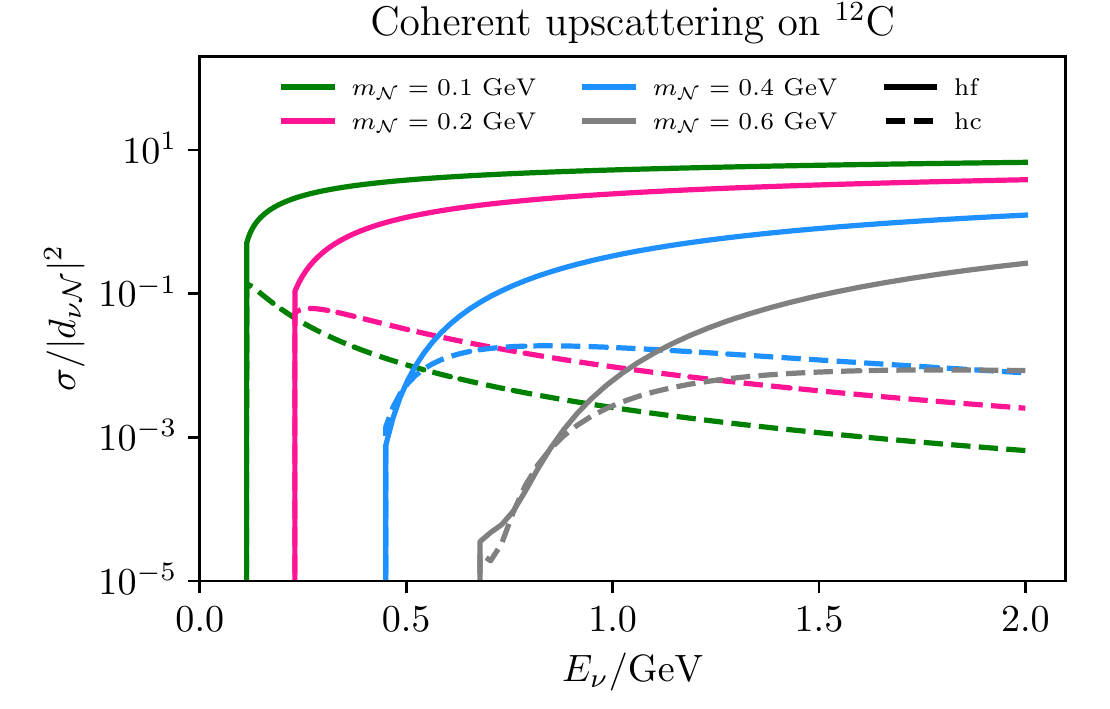}
    \includegraphics[width=0.49\textwidth]{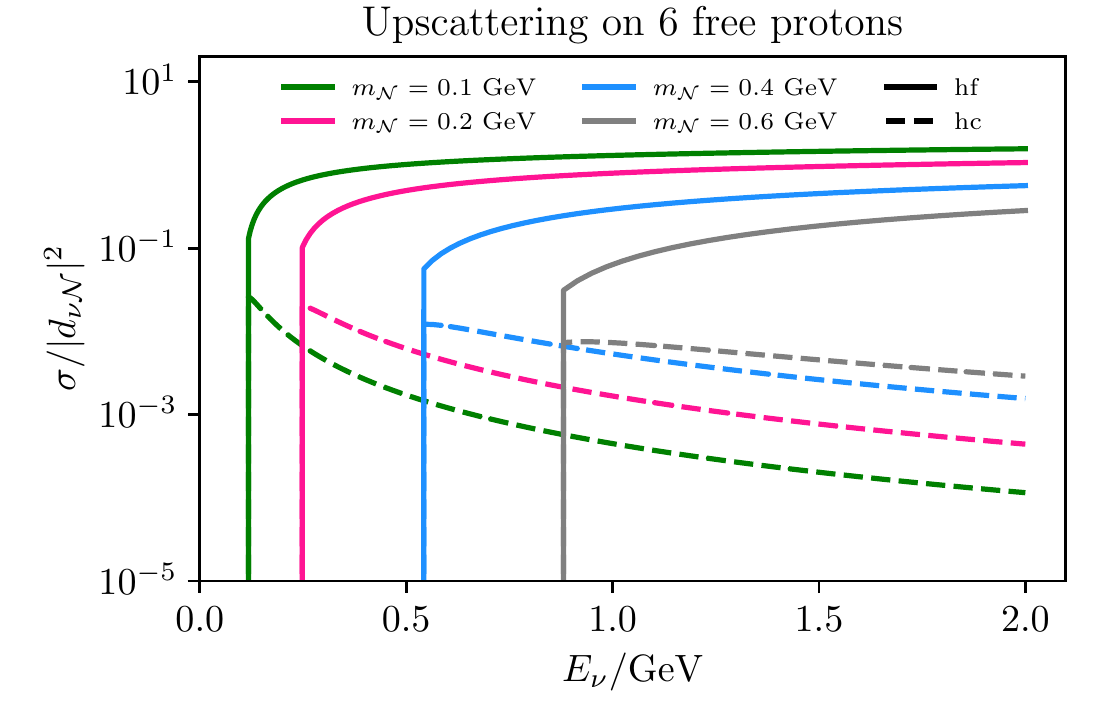}
    \caption{Comparison between helicity-flipping and helicity-conserving cross sections for coherent neutrino upscattering on Carbon (left) and free protons (right) for multiple values of the HNL mass $m_\mathcal{N}$.\label{fig:ups_xsecs}}
\end{figure*}

We have calculated both terms using the \texttt{DarkNews} code~\cite{DarkNews}, and show our results in \Cref{fig:ups_xsecs}. We show a comparison of the upscattering cross section on Carbon-12 for a few choices of HNL masses for both coherent and proton-elastic scattering regimes. Scattering on neutrons proceeds only via the neutron magnetic moment and is much smaller. We only include the proton-elastic contribution for MiniBooNE, where the proton would be invisible. This is a conservative approach when deriving the \minerva limits.

\begin{figure}[t]
    \centering
    \includegraphics[width=0.5\textwidth]{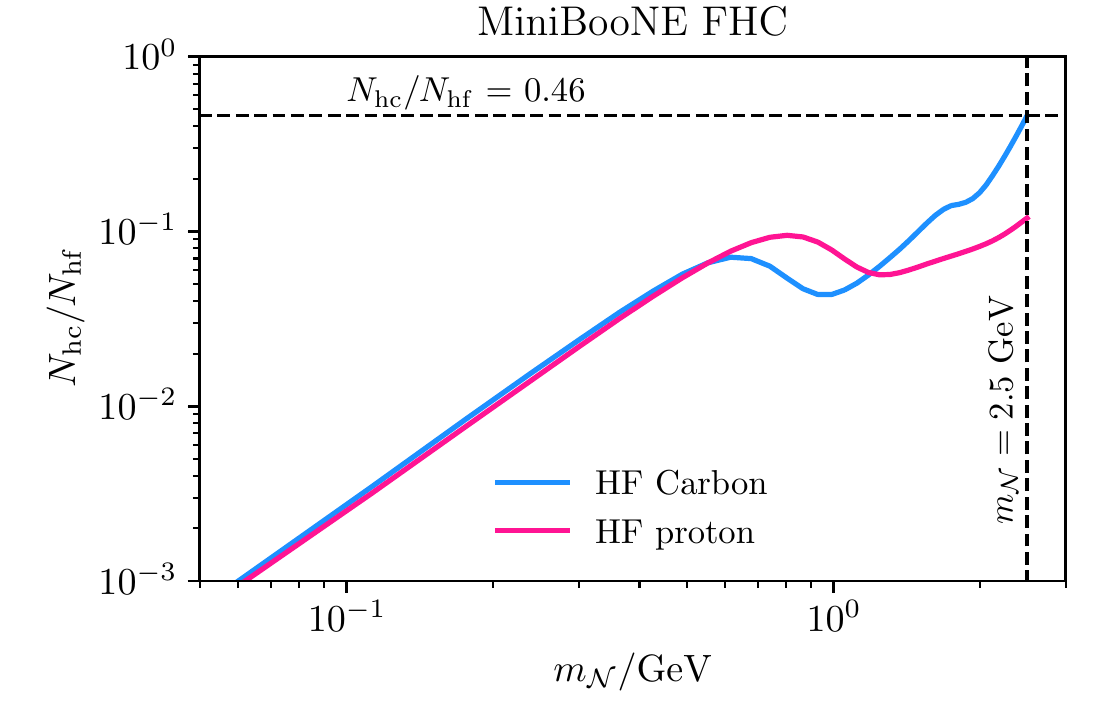}
    \caption{The ratio between helicity-conserving and helicity-flipping upscattering events on Carbon at MiniBooNE. \label{fig:helicity_ratio_MiniBooNE}}
\end{figure}

We also show the ratio between helicity-flipping and helicity-conserving upscattering events inside the MiniBooNE detector as a function of $m_{\mathcal{N}}$ in \Cref{fig:helicity_ratio_MiniBooNE}. The helicity-conserving part is a small correction, except at the very largest HNL masses, where the rate is significantly smaller due to the large energy threshold for upscattering.

\begin{figure*}
    \centering
    \includegraphics[width=0.8\textwidth]{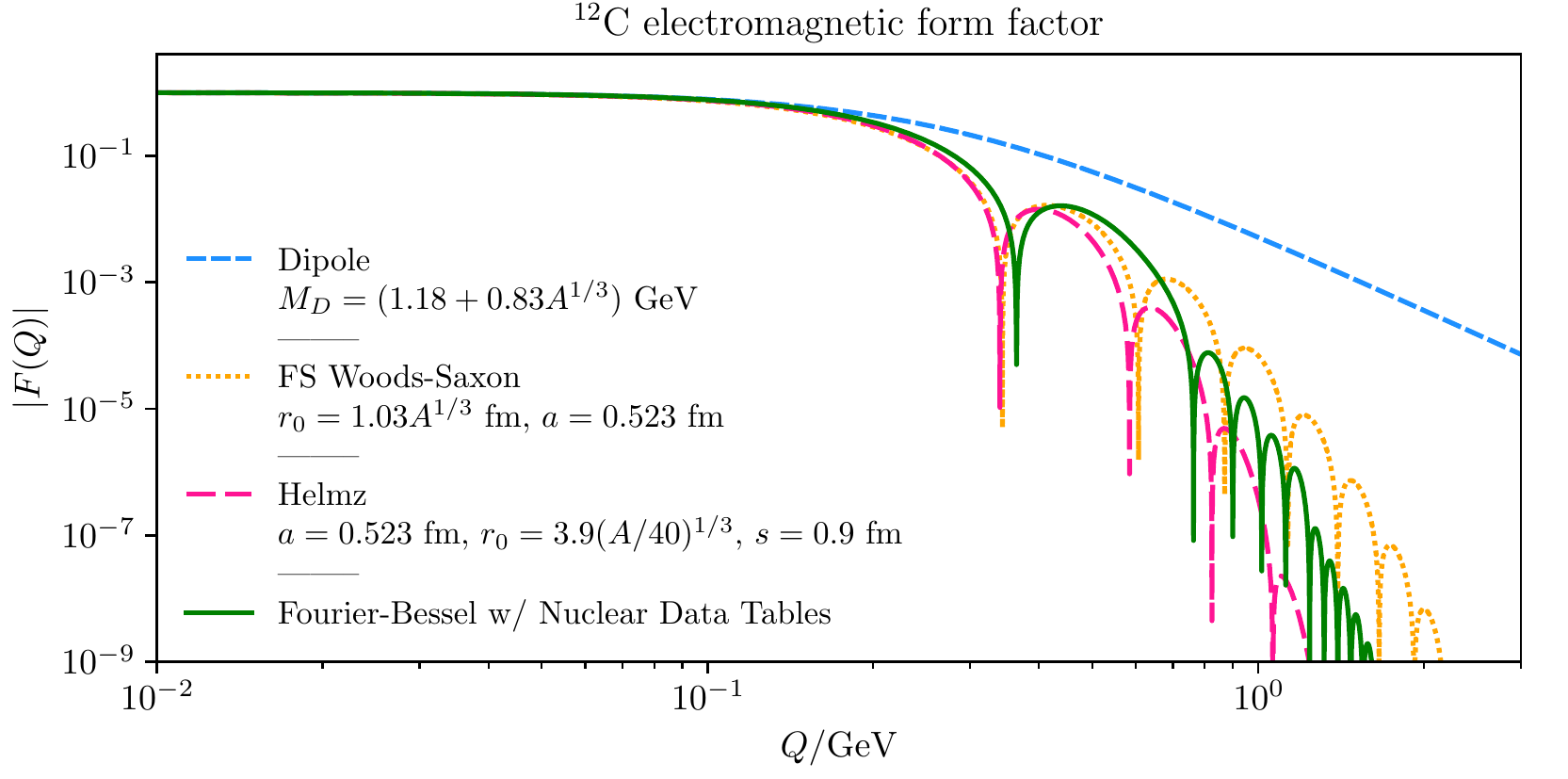}
    \caption{Comparison of the different nuclear form factors for $^{12}$C commonly used in the literature. The dipole form factor (dashed blue) significantly overestimates the form factor at large values of the momentum exchange $Q$. In this paper, we use the Fourier-Bessel parametrization (solid green) for nuclei for which nuclear data is available, otherwise, we implement the Fermi-Symmetrized (FS) Woods-Saxon parametrization (dotted yellow). \label{fig:form_factors}}
\end{figure*}

Several model-independent nuclear form factor parametrizations can be found in the literature. One of them is the Fourier-Bessel parametrization, which models the charge density in the nucleus as a series of Bessel functions with a radial cut-off of $R$.
\begin{equation}
    F^{\rm FB}(Q) = N\times \frac{\sin(Q R)}{QR} \sum_n \frac{(-1)^n a_n}{n^2 \pi^2 - Q^2},
\end{equation}
where $1/N = \sum_n (-1)^n a_n/n^2 \pi^2$ is a normalization factor, ensuring $F(0)=1$. The coefficients $a_n$ can be obtained from experimental data, which is available for a series of common nuclei~\cite{Fricke:1995zz,DeVries:1987atn,DeJager:1974liz}. We make use of the machine-readable files provided by Ref.~\cite{VT_NDT}.

For nuclei where the nuclear data cannot be found, we implement a Fermi-symmetrized Woods-Saxon form factor,
\begin{widetext}
\begin{equation}
    F^{\rm FS-WS} (Q) = \frac{3 \pi a}{r_0^2 + \pi^2 a^2} \frac{a \pi \text{cotanh}({\pi Q a})\sin({Q r_0})  - r_0 \cos({Q r_0})}{Q r_0 \sinh{(\pi Q a)}},
\end{equation}    
\end{widetext}
where $a=0.523$~fm, $r_0=1.03 \times A^{1/3}$~fm. These form factors correctly describe the finite nuclear radius and lead to a strong suppression of coherent scattering for $Q\gtrsim 200$~GeV. They should be contrasted with the simpler dipole parametrization
\begin{equation}
    F^{\rm Dip}(Q) = \frac{1}{1 + \frac{Q^2}{M_D^2}}.
\end{equation}
used in Ref.~\cite{Vergani:2021tgc}, with $M_D = 1.18 + 0.83* A^{1/3}$, and the Helmz form factor
\begin{equation}
    F^{\rm Helmz}(Q)  \frac{3 |j_1(Q R)|}{QR} e^{-Q^2 s^2/2},
\end{equation}
with $a = 0.523$~fm, $s = 0.9$~fm, and $R = 3.9$~fm. 

We provide a comparison of the aforementioned nuclear form factors in \Cref{fig:form_factors}. 
It is evident that the dipole parametrization overestimates the cross section at large values of momentum exchange $Q$. 
The more sophisticated form factors used in this work produce more forward angular distributions at MiniBooNE than what was found in the previous study of Ref.~\cite{Vergani:2021tgc}.

\bibliographystyle{apsrev}
\bibliography{neutrissimo}
\end{document}